\documentclass[11pt,preprint]{aastex}

 \newcommand{\feii}{\ion{Fe}{2}}
 \newcommand{\kms}{km\,s$^{-1}$}
 \newcommand{\lb}{$\lambda$}
 \newcommand{\hbeta}{H$\beta$}

 \slugcomment{Accepted to ApJ.}

 \shorttitle{Infrared Fe\,II Emission in NLS1 Galaxies}
 \shortauthors{Rodr\'{\i}guez-Ardila et al.}

\begin{document}

 \title{Infrared \feii\ Emission in Narrow-Line Seyfert 1 Galaxies}

 \author{A. Rodr\'{\i}guez-Ardila\altaffilmark{1} and S. M. Viegas}
 \affil{Instituto Astron\^omico e Geof\'{\i}sico - Universidade de S\~ao
 Paulo,
 Av. Miguel Stefano 4200, CEP 04301-904, S\~ao Paulo, SP, Brazil}

 \author{M. G. Pastoriza}
 \affil{Departamento de Astronomia - UFRGS. Av. Bento Gon\c calves 9500,
 CEP 91501-970, Porto Alegre, RS, Brazil}

 \and

 \author{L. Prato\altaffilmark{1}}
 \affil{Department of Physics and Astronomy, UCLA, Los Angeles, CA
 90095-1562}

 \altaffiltext{1}{Visiting Astronomer at the Infrared Telescope facility, which
is operated by the University of Hawaii under contract from the National 
Aeronautics and Space Administration}

 \begin{abstract}
 We obtained 0.8-2.4 $\mu$m spectra at a 
 resolution of 320 km\,s$^{-1}$ of four narrow-line Seyfert
 1 galaxies in order to study the near-infrared properties of 
 these objects. We focus on the analysis of the \feii\ emission
 in that region and the kinematics of the low-ionization broad lines.
 We show that the 1$\mu$m \feii\ lines (\lb9997, 
 \lb10501, \lb10863 and \lb11126) are the strongest \feii\
 lines in the observed interval. 
 For the first time, primary cascade lines of \feii\ arising from 
 the decay of upper levels pumped by Ly$\alpha$ fluorescence are  
 resolved and identified in active galactic nuclei.
 Excitation mechanisms leading to the emission of the 1$\mu$m \feii\
 features are discussed. A combination of Ly$\alpha$ fluorescence
 and collisional excitation are found to be the main contributors. 
 The flux ratio between near-IR \feii\ 
 lines varies from object to object, in contrast to what 
 is observed in the optical region. A good correlation
 between the 1$\mu$m and optical \feii\
 emission is found. This suggests that the upper $z^{\rm 4}F^{\rm 0}$ and 
 $z^{\rm 4}D^{\rm 0}$ levels from which the bulk of the optical
 \feii\ lines descend are mainly populated by the transitions leading
 to the 1$\mu$m lines. The width and profile shape of \feii\ \lb11127, 
 \ion{Ca}{2} \lb8642 and \ion{O}{1} \lb8446 are very similar but 
 significantly narrower than Pa$\beta$, giving
 strong observational support to the hypothesis that the 
 region where \feii, \ion{Ca}{2} and \ion{O}{1}
 are produced are co-spatial, interrelated kinematically and most 
 probably located in the outermost portion of the BLR.

 \end{abstract}


 \section{Introduction}

 Narrow-line Seyfert 1 galaxies (hereafter NLS1s) are a 
 peculiar group of Seyfert 1 objects, recognized by 
 \citet{op85} for their optical properties: 
 (a) the broad components of the permitted lines have FWHM $<$ 
 2000 \kms, narrower that those of the classical Seyfert 
 1s (i.e $\sim$4000 \kms), (b) the [\ion{O}{3}]/\hbeta\ ratio is 
 less than 3, and (c) usually, strong \feii\ emission dominates 
 the optical and ultraviolet spectrum. In the 
 soft and hard X-ray bands NLS1s also seem to have 
 peculiar properties. In most cases they are 
 characterized by a very steep spectrum and extreme 
 and rapid variability \citep{bbf96, ley99}. Up to now, 
 no single model can explain all of the observed properties 
 of these objects and much effort has recently been made 
 in order to understand their central engines. Variability 
 studies \citep{pet2000} point out that the ultimate cause 
 for their peculiar behavior may be attributable to a smaller black 
 hole mass ($\sim 10^{6}$ M$\odot$) together with a  
 near- or super-Eddington accretion rate.

 NLS1 galaxies are ideal objects for studying many features 
 which are clearly separated in their spectra but are 
 heavily blended in classical Seyfert 1s because of the broadness 
 of their permitted lines. Such is the case with the \feii\ 
 emission lines, which produce a ubiquitous pseudo-continuum 
 from the ultraviolet through the near-infrared (near-IR).
 This pseudo-continuum 
 carries a large amount of energy from the BLR
 and is not yet well
 understood. In the optical and UV regions, the \feii\
 lines have been well studied \citep{wnw85, jol91, lip93}. 
 The main processes believed to contribute to the bulk of that
 emission are continuum fluorescence
 via UV resonance lines, self-fluorescence via overlapping
 \feii\ transitions and collisional excitation. Nonetheless,
 most models that include these processes cannot account 
 for the observed strength of the \feii\ lines. \citet{pen87},
 \citet{sp98} and \citet{ver99} considered that fluorescent 
 excitation by Ly$\alpha$ may be important
 in understanding the \feii\ 
 spectrum in active galactic nuclei (AGNs), from the 
 ultraviolet to the IR. Hovewer, the lack of suitable 
 data in the latter region has hindered tests of
 predictions. 
  
 In the near-IR region, \feii\ emission lines have been mostly
 observed in a variety of Galactic emission-line stars
 \citep{hp89, rudy91, ham94}.  At least in these sources,
 the \feii\ lines redward of 7000 \AA\ can be attributed
 to cascades from excited states which are pumped by
 resonant absorption of \ion{H}{1} Ly$\alpha$. In all
 cases, the strongest \feii\ lines are the ones located
 at $\lambda$9997, $\lambda$10171, $\lambda$10490,
 $\lambda$10501, $\lambda$10863 and $\lambda$11126,
 emitted after the decay of the common upper term ($b~^{\rm 4}G$). 
 Because of their close proximity in wavelength, they are termed
 the 1$\mu$m \feii\ lines. Other important \feii\ lines
 also detected in Galactic sources are located in
 the 8500$-$9300 \AA\ interval. They are primary cascade lines
 descending from the upper 5p levels to the lower
 $e^{\rm 4}D$ and $e^{\rm 6}D$ terms. With the advent of 
 a new generation of IR detectors with greater sensitivity 
 and resolution in this region, it is now possible to 
 search for these lines in AGNs, mainly in NLS1s, and
 test the validity of the models of \feii\ emission. This 
 is a crucial step in understanding the physical processes 
 at work in the central engines of these objects.

 The only detection to date of the 1$\mu$m 
 \feii\ lines in an AGN was reported by 
 \citet[hereafter RMPH]{rudy00} in their 0.8 -- 2.5$\mu$m 
 spectrophotometric observation of I\,Zw\,1, the 
 archetypical NLS1 galaxy. Their data clearly reveal \feii\
 $\lambda$9997, $\lambda$10501, $\lambda$10863 and 
 $\lambda$11126. Based on the absence of the crucial cascade 
 lines that feed the common upper state where the
 1$\mu$m \feii\ lines originate (assuming Ly$\alpha$ fluoresence as the
 dominant mechanism), as well as the relatively low energy of that
 state, RMPH suggest that the observed lines are collisionally 
 excited. No individual identification of \feii\ lines
 in the 8500$-$9300 interval has yet been done in Seyferts. 
 But it is believed that the strong broad feature, centered at
 \lb9220 \AA\ and observed in many AGNs is, in part,
 due to \feii\ \citep{morward89}.
 
 In this paper we present mid-resolution ($\sim$ 320 km\,s$^{-1}$),
 near-IR spectroscopy of four NLS1 galaxies (1H1934-063, Ark\,564, 
 Mrk\,335, and Mrk\,1044), not previously observed in
 this spectral region, whose optical spectra reveal
 strong to moderate \feii\ emission. In \S 2 we present
 the observations, data reduction, and a brief description of
 the most conspicuous spectroscopic features observed in each
 galaxy. The excitation mechanism
 of the \feii\ lines and the observed relative intensity are
 discussed in \S 3. The kinematics of the BLR derived from
 the low-ionization lines is treated in \S 4. Conclusions
 are presented in \S 5.

 \section{Observations, Data Reduction and Results} \label{observa}

 \subsection{Observations}

 The data were obtained at the NASA 3m Infrared Telescope
 Facility (IRTF) on 2000 October 11 and 13 (UT) 
 with the SpeX facility spectrometer \citep{rayner98, rayner01}.  
 In the short wavelength
 cross-dispersed mode, SpeX provides nearly continuous coverage
 from 8200 \AA~ through 24000 \AA.  Ten minute integrations,
 nodding in an off-on-on-off source pattern, were combined to
 produce final spectra.  Typical total integration times ranged
 from 30$-$40 minutes.  A0V stars were observed
 near targets to provide telluric standards at similar airmasses.
 Seeing was $\sim$1.$''$0.  An 0.$''$8 slit yielded an
 instrumental resolution of 320 \kms. 

 \subsection{Data Reduction}

 The spectral extraction and wavelength calibration procedures were
 carried out using Spextool, the in-house software developed 
 and provided by the SpeX team for the IRTF 
 community \citep{cvr01}\footnote{Spextool is available from the
 IRTF website: http://irtf.ifa.hawaii.edu/Facility/spex/spex.html}.
 Each on-off source pair of observations was reduced individually
 and the results summed to provide a final spectrum.  A 1$''$ or
 3$''$ (in the case of 1H\,1934-063) aperture was used to extract the
 spectra. Observations of an argon arc lamp enabled wavelength calibration
 of the data; the RMS of the dispersion solution for the four target
 galaxies was $\sim$2$\times$10$^{-5}\mu$m.

 Wavelength calibrated final target spectra were divided
 by AOV stars observed at a similar airmasses.  However, even small
 differences in airmass between the target galaxies
 and standard stars resulted in residuals, primarily evident in the
 spectrum of Mrk 1044.  The spectra were each multiplied
 by a black body function corresponding to 10,000 K in order to
 restore the true continuum shape of the targets.

 Spectra were flux calibrated by normalizing to the $K$ band
 magnitude, 6.303 mag, of one of the observed A0V standard stars, 
 HD\,182761. We used the task STANDARD of the ONESPEC package of 
 IRAF\footnote{IRAF is distributed by NOAO, which is operated
 by AURA Inc., under contract to the NSF.} for this 
 purpose. In this process, a blackbody flux 
 distribution based on the magnitude and spectral
 type of the star was employed. The error in the flux calibration 
 is $\sim$10\%, estimated from comparison with an overlapping 
 visual wavelength, red spectrum of 1H1934-063 at 0.8$-$0.95 $\mu$m 
 and previously published IR spectrophotometry for 
 MRK\,335 \citep{rudy82}. Nonetheless, this calibration is
 not absolute. For this reason, the uncertainties quoted in this paper
 reflects the errors derived from the S/N of the spectra. 

 We correct the
 spectra for Galactic extinction, as determined from the 
 {\it COBE/IRAS} infrared maps of \citet{sfd98}.
 The value of the Galactic {\it E(B-V)} as well as basic 
 information for each object is listed in 
 Table~\ref{tabledata}. Finally, each spectrum
 was shifted to rest wavelength. The value of $z$ adopted
 was determined by averaging the redshift measured from the 
 strongest lines, usually \ion{O}{1} \lb8446, [\ion{S}{3}]
 \lb9531, \feii\ \lb9997, Pa$\delta$, \ion{He}{1} \lb10830,
 \ion{O}{1} \lb11287, Pa$\beta$, Pa$\alpha$ and Br$\gamma$.
 In all cases, the radial velocities were in very good 
 agreement with the values reported in the literature.

 \subsection{Results}

\begin{figure}
 \plotone{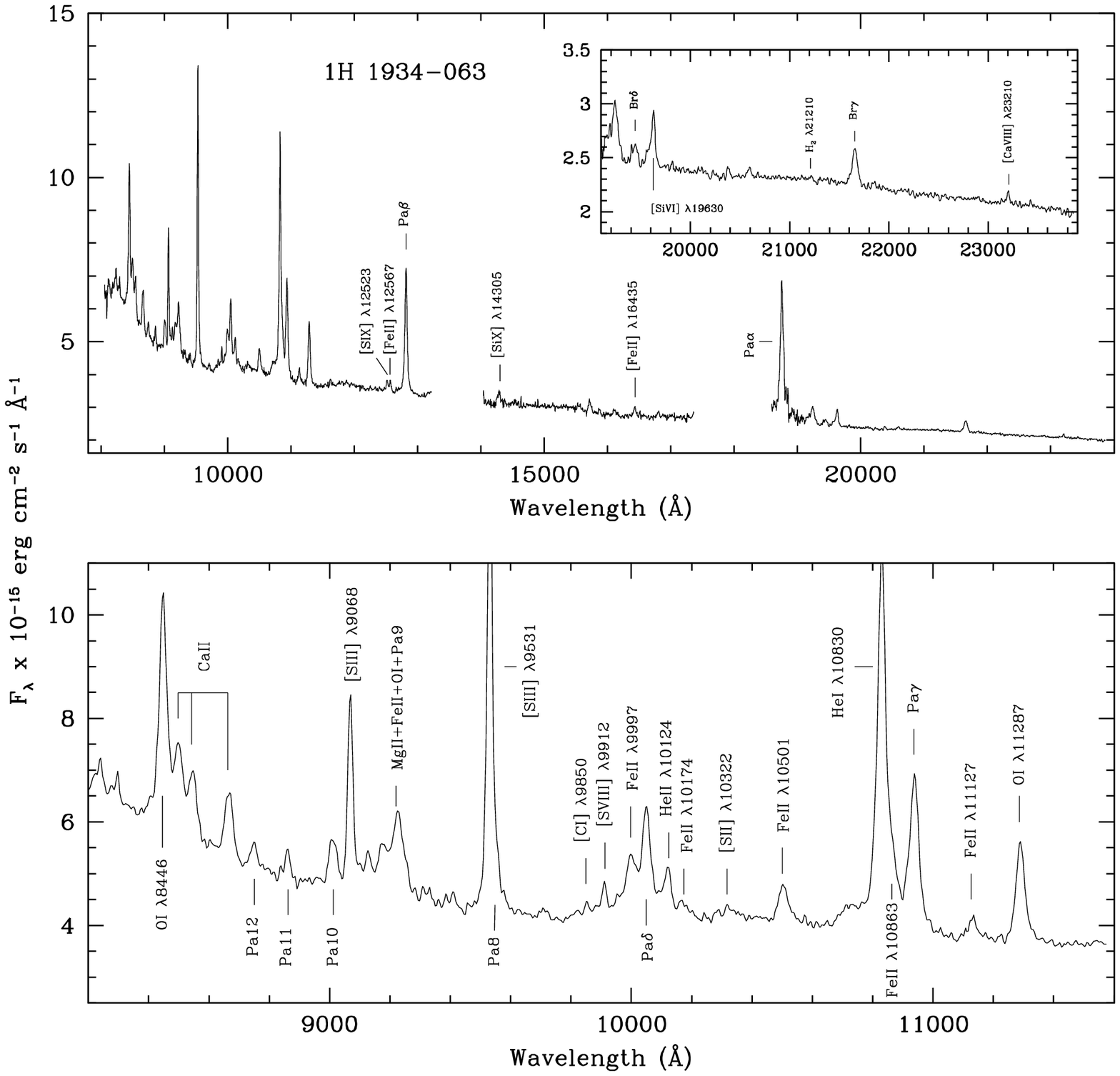}
 \caption{Near-IR spectrum of the NLS1 galaxy 1H\,1934-063 in rest
 wavelengths. The upper panel shows the spectrum in the wavelength
 interval 0.8--2.4$\mu$m, and the box in the upper right a zoom of 
 the K-band spectrum, with the identification of the most important 
 emission lines. The lower panel displays the spectrum around 
 1$\mu$m and identifies the \feii\ and other permitted and forbidden 
 lines present in this region. \label{fig1}}
 \end{figure}

\begin{figure}
 \plotone{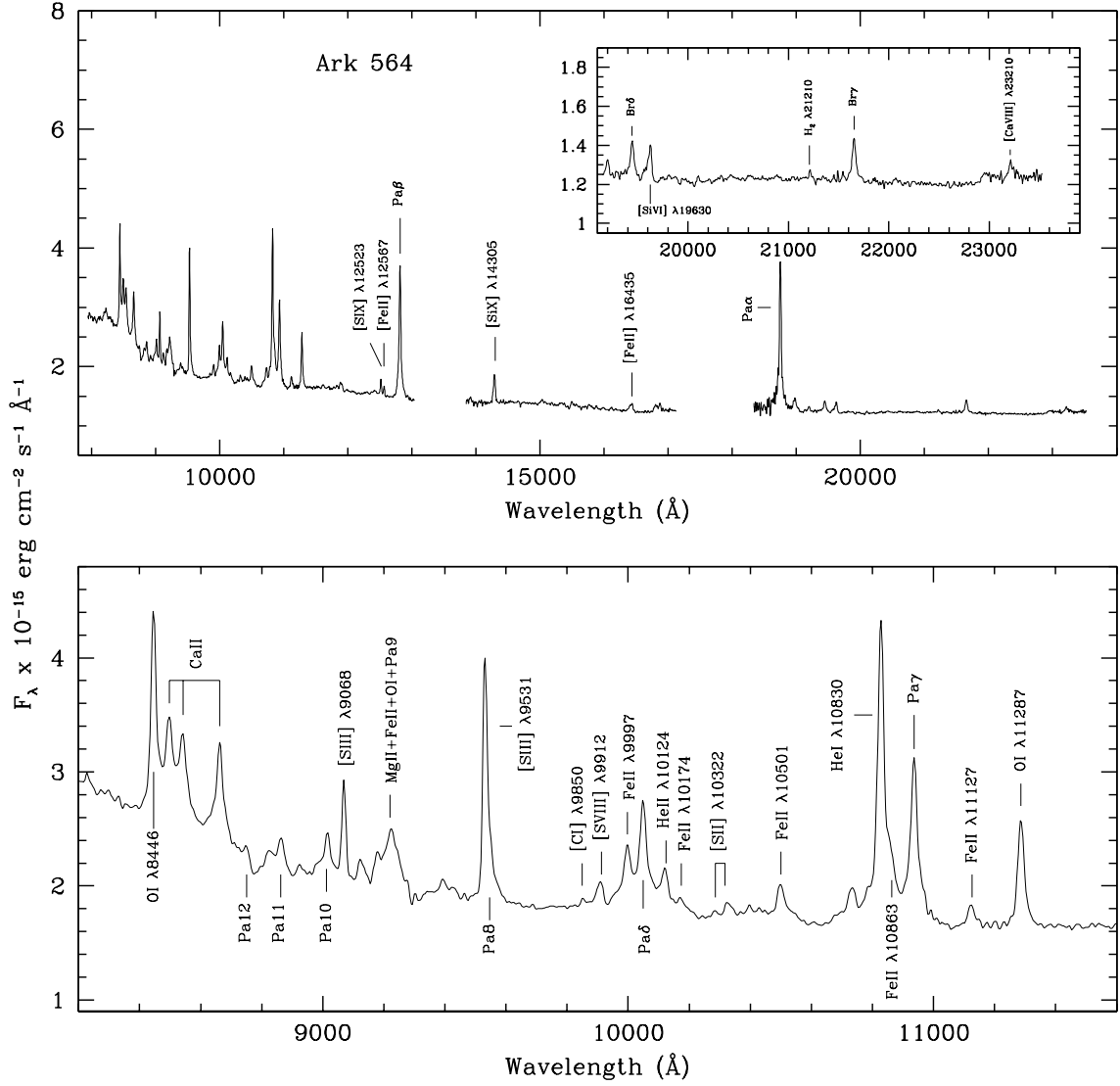}
 \caption{The same as Figure~\ref{fig1} for Ark\,564. \label{fig2}}
 \end{figure}

\begin{figure}
 \plotone{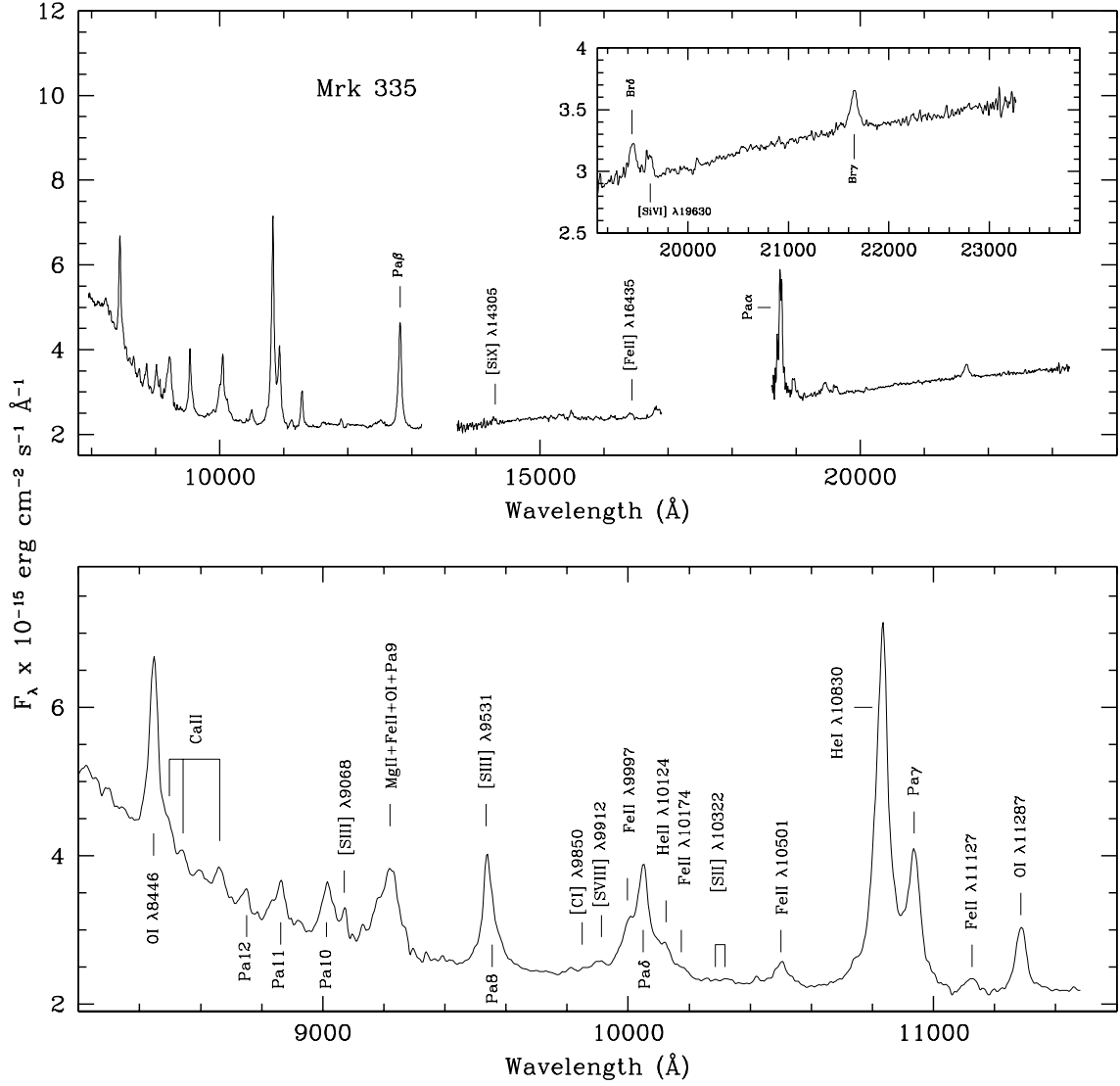}
 \caption{The same as Figure~\ref{fig1} for Mrk\,335. \label{fig3}}
 \end{figure}

\begin{figure}
 \plotone{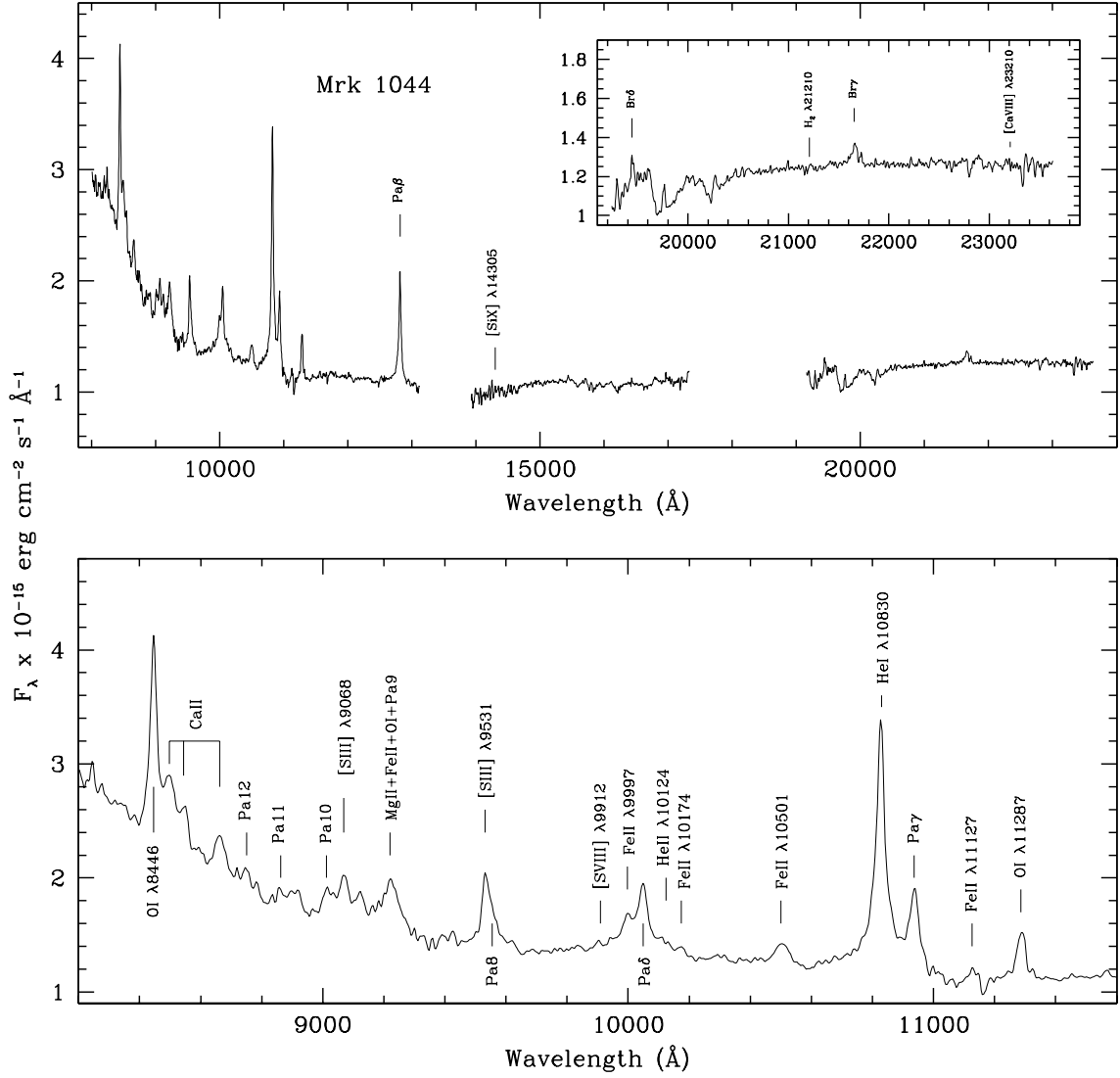}
 \caption{The same as Figure~\ref{fig1} for Mrk\,1044. \label{fig4}}
 \end{figure}

 The resulting extracted spectra in the wavelength 
 interval 0.8 -- 2.4$\mu$m, corrected for
 redshift, are displayed in Figures~\ref{fig1} -- \ref{fig4}.  
 Identification of the known emission lines are 
 given for each spectrum. Prominent lines of \feii,
 \ion{H}{1}, \ion{He}{1}, \ion{O}{1} and
 \ion{Ca}{2} are observed in the four galaxies,
 similar to those present in I\,Zw\,1. In addition, 
 \ion{He}{2} \lb10124, as well as lines of
 [\ion{S}{3}] $\lambda\lambda$9070, 9531, 
 [\ion{S}{8}] \lb9912, [\ion{S}{9}] \lb12523, 
 [\ion{Fe}{2}] \lb12567, [\ion{Si}{10}] \lb14305 
 and [\ion{Si}{6}] \lb19630, not present in I\,Zw\,1,
 are clearly detected in our data. A 
 detailed analysis of the narrow emission lines is 
 beyond the scope of this paper and will be discussed 
 in a future publication \citep{roar01a}. 

 In order to measure the flux of the emission lines
 we have assumed that they can be represented by a
 single or a sum of Gaussian profiles. The 
 LINER routine (Pogge \& Owen 1993), a $\chi^{2}$
 minimization algorithm that fits as many as eight
 Gaussians to a profile, was
 used for this purpose. Table~\ref{tableflux} list 
 fluxes and FWHM for the detected
 \feii\ and other permitted lines. The FWHM values were
 corrected for instrumental broadening ($\sim$ 320 \kms).
 Except for \feii\ \lb10501\ and
 \lb11127, which are isolated features, the 
 \feii\ lines are blendend with \ion{He}{1} 
 $\lambda$10830 plus Pa$\gamma$ (\feii\ $\lambda$10863) 
 or \ion{He}{2} $\lambda$10124 plus Pa$\delta$ 
 (\feii\ $\lambda$9956, $\lambda$9997 
 and $\lambda$10171). For consistency, in these latter
 cases the \feii\ and Paschen lines were constrained to 
 have the same width as those derived for \feii\ $\lambda$11127
 and Pa$\beta$, respectively. Nonetheless, it is interesting
 to note that even if no constraint was imposed, the
 resulting line widths were in agreement with those found
 for the isolated lines. 

 The above procedure also allowed us to separate the
 contribution of the NLR in those lines that are emitted 
 both in the BLR and the NLR. The results show that 
 in the Paschen lines, for instance, nearly half the total 
 flux is emitted by the NLR, very similar to what was found
 by \citet{ro00} for the Balmer lines. Table~\ref{tablefluxbn}
 list the fluxes and FWHM of the Pashen  and \ion{He}{1}
 \lb10839 lines that could be deblended into a narrow and
 broad component. All the values of FWHM were corrected for
 instrumental broading. In Mrk\,1044 it was not possible to find
 a consistent solution for Pa$\delta$, Pa$\gamma$ and \ion{He}{1}
 \lb10839 due to the lower S/N of the spectrum. For this
 reason, the total fluxes associated with each of the latter three
 lines are reported but must be understood as the sum of
 the contribution of both the NLR and BLR.

\begin{figure}
 \plotone{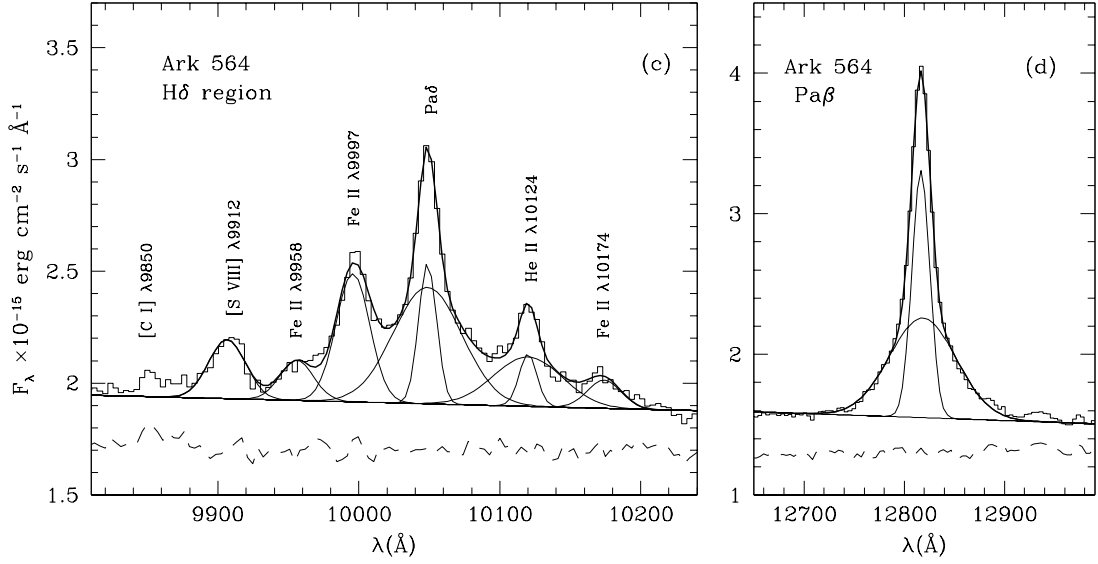}
 \caption{Example of the deblending procedure applied to Pa$\beta$ and
 Pa$\delta$ for Ark\,564. The histogram is the data and the thick
 line the fit to the observed blend. Individual Gaussian components are
 in thin lines. The residuals of the fits are drawn in dashed
 lines.\label{deblend}}
 \end{figure}

 The presence of emission from the NLR is demonstrated by
 the observation of strong [\ion{S}{3}] \lb9068, \lb9531.
 In order to test the validity of the Gaussian 
 approximation, we used the technique applied by
 \citet{ro00} to separate the narrow and broad flux in 
 Pa$\beta$, consisting of using
 the profile of a line known to be emitted exclusively in 
 the NLR as a template. In the current work, we assumed that the width
 of the narrow component of the permitted lines would be 
 similar to that of [\ion{S}{3}] \lb9531, the strongest
 forbidden line observed.  Based on this assumption, the
 [\ion{S}{3}] \lb9531 line was scaled to the peak
 intensity of a given permitted line and then subtracted off
 in different proportions until the residuals consist 
 of a pure broad component with no absorption superimposed.
 This procedure has the advantage of being free from
 any assumption regarding the form of the line profiles.
 The results were almost identical 
 to those obtained using the Gaussian approximation. 
 Figure~\ref{deblend} shows examples of the deblending
 of Pa$\beta$ and the region around Pa$\delta$ for 
 Ark\,564.

 In the rest of this section we will give 
 brief descriptions of the most important features
 observed in the NLS1 spectra.

 \subsubsection{1H\,1934-063}

 1H\,1934-063 is a relatively poorly known NLS1
 but an excellent laboratory to study many emission
 line processes. Its optical spectrum, described by
 \citet{ro00}, is full of both permitted and forbidden
 lines. High ionization lines such as [\ion{Fe}{10}]
 and [\ion{Fe}{14}] are very prominent. In the
 near-IR it also displays a very 
 rich spectrum, both in permitted and forbidden lines 
 (see Figure~\ref{fig1}). \ion{He}{1} \lb10830, 
 Pa$\alpha$, Pa$\beta$, Pa$\delta$, \ion{O}{1} \lb8446, 
 \lb11287, the \ion{Ca}{2} triplet in emission and
 the 1$\mu$m \feii\ lines are strong. In addition,
 Br$\gamma$ and Br$\delta$ are clearly observed. A very
 conspicous blend of low-ionization species, centered
 in 9220 \AA\ is present. We identified \feii\ \lb9177, \lb9202
and \lb9256 as well as Pa9 \lb9227 as contributors to this
feature. \ion{Mg}{2} \lb9218, \lb9244 are probably present
but heavily blended with the Pa9. The continuum emission is very steep, 
 and follows a simple power-law form.
 High-ionization, forbidden lines of [\ion{S}{8}],
 [\ion{S}{9}], [\ion{Si}{10}], [\ion{Si}{7}] and
 [\ion{Ca}{8}] are clearly visible as well as a
 blip at the expected position of the 
 H$_{2}$ (1,0)S(1) \lb2.122$\mu$m.  [\ion{S}{3}]
 \lb9531 is, by far, the strongest forbidden line
 detected.

 \subsubsection{Ark\,564}

 Ark\,564 is a widely known and well studied NLS1 galaxy,
 mainly in the UV and X-ray region \citep{bbf96, cren99, 
 com00, ball00}. It is the brightest NLS1 in the 2-10 keV range; the
 {\it ROSAT} and {\it ASCA} observations reveal a complex
 X-ray spectrum \citep{vaug99}. In the near-IR, this
 object is also very rich in emission lines 
 (see Figure~\ref{fig2}), similar 
 in intensity to those of 1H\,1934-063. The permitted lines
are rather narrow, just slightly broader than the forbidden lines.
The 1$\mu$m lines
 are strong and well resolved. Br$\gamma$ is clearly
 detected as well as the molecular H$_{2}$ line (1,0)S(1) 
 at \lb2.121$\mu$m. The \ion{He}{1} \lb10830,
 \ion{Ca}{2} triplet,
 and \ion{O}{1} \lb8446, \lb11287
 lines are all strong. The \lb9220 blend is 
 present with \feii\ \lb9177, \lb9202, \lb9256, Pa9 and 
\ion{Mg}{2} \lb9244 clearly resolved. The forbidden line 
spectrum is remarkable,
 displaying strong low and high ionization lines such
 as [\ion{S}{3}] \lb9068,9031, [\ion{C}{1}] \lb9850,
 [\ion{S}{8}] \lb9912, [\ion{S}{2}] \lb10286,10322,
 [\ion{S}{9}] \lb12523, [\ion{Si}{10}] \lb14305 and
 [\ion{Si}{6}] \lb19630. The continuum emission rises 
 steeply from Pa$\beta$ to shorter wavelengths and is flat 
 between 1.5 and 2.6$\mu$m.

 \subsubsection{Mrk\,335}

 This object is a strong UV and X-ray source
 \citep{bbf96, cren99, com00, ball00}. The near-IR
 spectrum is dominated by \ion{He}{1} \lb10830
 and the \ion{H}{1} Paschen lines (Figure~\ref{fig3}).
 The 1$\mu$m \feii\ feature is not as strong as in the former two 
 galaxies. In addition, \feii\ \lb11127 and Pa$\alpha$ are 
 affected by atmospheric absorption. The \ion{Ca}{2} 
 triplet in emission, although present, is weak.
 The \lb9220 feature is rather strong. From the Gaussian fitting, 
we identified \feii\ \lb9177, \lb9202, \lb9256 and Pa9
as individual components of this blend. The continuum 
 emission of this galaxy is particularly interesting:
 there is a broad minimum around 1.4$\mu$m and a rapid
 rise towards longer wavelengths. It
 probably represents the shift from a non-thermal
 continuum to the thermal dust emission regime.
 In their near-IR spectrophotometry
 of Mrk\,335, \citet{rudy82} pointed out that the colours of this
 object are among the reddest and most nonstellar
 reported for any Seyfert Galaxy. They also conclude
 that the small flux at 10$\mu$m and 20$\mu$m indicates that
 Mrk\,335 does not contain large amounts of cool dust
 and that the bulk of near-IR luminosity is produced
 by a relatively small amount of warm dust. 

 \subsubsection{Mrk\,1044} \label{mrk1044}

 The dominant feature in the spectrum
 of this galaxy is \ion{He}{1} \lb10830 
 (Figure~\ref{fig4}). The 1$\mu$m 
 \feii\ lines are present but only 
 \lb9997 and \lb10501 are easily identified. \feii\ 
 \lb11127 is severly affected by atmospheric absorption
 so it is only possible to derive a lower limit for
 the flux of this line. \ion{O}{1} \lb8446, \lb11287,
 \ion{Ca}{2} \lb8498, \lb8542, \lb8662 and the \lb9220
 feature are other important signatures of the BLR visible
 in this spectrum.  The NLR contribution, as in
 Mrk\,335, is rather weak, with only [\ion{S}{3}] \lb9531
 clearly observed. The continuum emission also shows
 a broad minimum around 1.4$\mu$m. It is very
 steep blueward and mostly flat redward of 1.4$\mu$m.

 \section{Discussion}

 The spectra presented here not only confirm the
 presence of the 1$\mu$m \feii\ lines in NLS1s, but also
 that they are the strongest and most prominent
 \feii\ lines observed in the 0.8$-$2.4 $\mu$m interval. 
 This suggests that the 1$\mu$m \feii\ lines 
 are a common feature in these objects, similar to
 the \feii\ bumps at both sides of H$\beta$  in the 
 optical region. In addition, fainter \feii\ lines 
 are clearly detected in the interval 8700$-$9300 \AA. 
 Our goal is to discuss the dominant excitation mechanism 
 for these lines and to confirm if, as in I\,Zw\,1, 
 collisional excitation is the most plausible one. For 
 this purpose, our data as well as information from other 
 wavebands will be used. It is also important to determine 
 if the relative intensities of the 1$\mu$m \feii\ lines 
 vary from object to object, or if, as in the optical region,
 they are approximately constant.

 \subsection{Excitation Mechanisms of the near-IR \feii\ lines}

 Excitation mechanisms invoked for the \feii\ emission
 include continuum fluorescence via the UV resonance lines,
 self-fluorescence via overlapping \feii\ transitions,
 and collisional excitation. The latter two are thought
 to contribute to the bulk of the \feii\ emission
 \citep{wnw85}. Nonetheless, most models that take into
 account these processes fail at reproducing the observed
 intensity of UV and optical \feii\ lines.
 
 \subsubsection{What we have learned from stars}

\begin{figure}
 \plotone{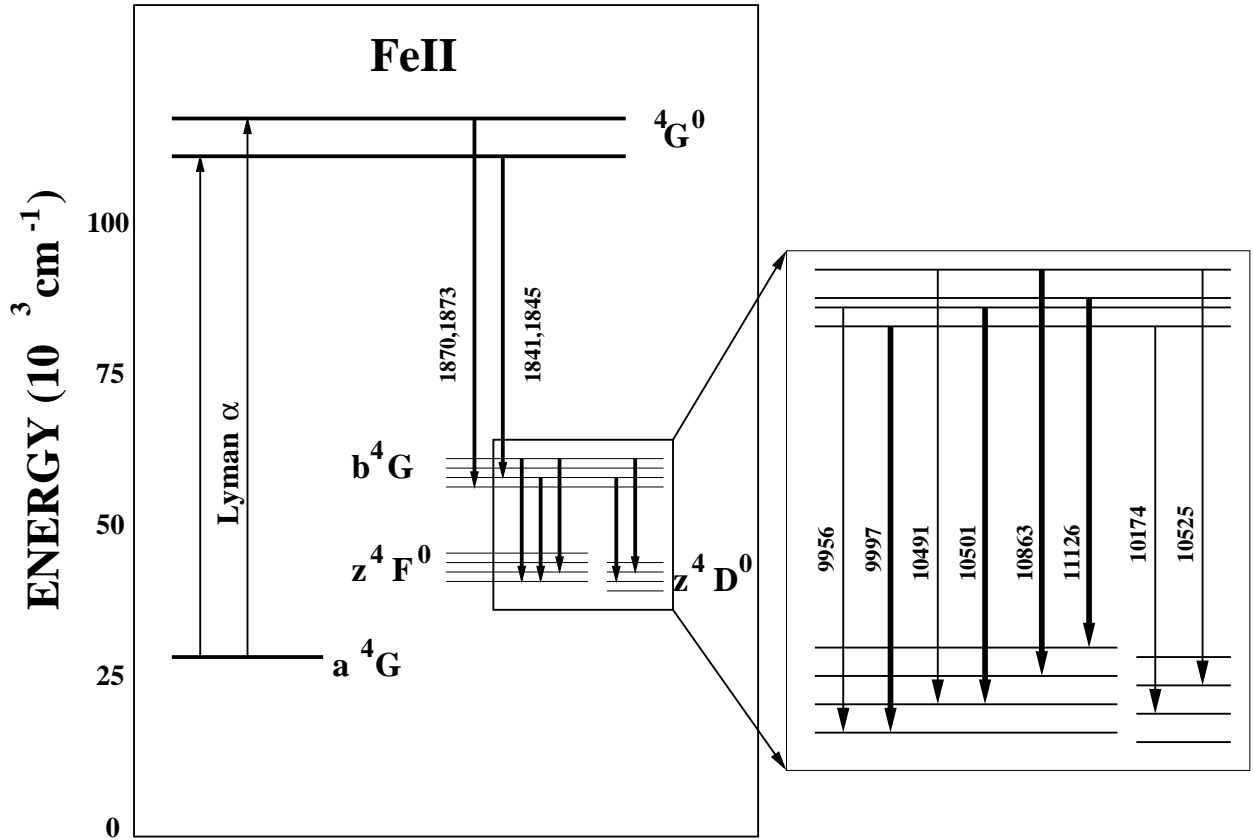}
 \caption{Energy level diagram for \feii\ illustrating the
 transitions to $z~^{4}F^{0}$ and $z~^{4}D^{0}$
 from the upper $b^{\rm 4}G$ level leading to the 1$\mu$m 
 lines. The strongest lines observed are drawn in bold face. The
 excitation via Ly$\alpha$ fluorescence is also shown. \label{levels}}
 \end{figure}

 Based on the observation of \feii\ lines in low resolution 
 {\it IUE} spectra of cool giants and supergiants, \citet{joh83} 
 and \citet{jojor84} proposed a large set of pumping channels
 for Ly$\alpha$ to excite \feii\ levels with
 excitation energy around 10 eV (5p levels). The decay route
 from these levels include $e^{\rm 4}D$ and $e^{\rm 6}D$ of 
 3$d^{\rm 6}(^{\rm 5}D)5s$, which lie at wavelengths 
 $\sim$8000$-$9600 \AA. These two levels
 decay giving rise to multiplets such as UV 399, 391, 380 ($\sim$
 2850\AA), 373 ($\sim$2770\AA) and 363 ($\sim$2530\AA), which supply
 approximately 20\% of the total energy in \feii\ lines
 between 2000 and 3000 \AA\ \citep{pen87}.
 In addition, \citet{jojor84} proposed Ly$\alpha$ 
 fluorescence to pump a further $^{\rm 4}G^{\rm 0}$ level at 
 $\sim$13 eV, whose primary decay to b$^{\rm 4}G$ produces the UV 
 lines at \lb1841, \lb1845, \lb1870 and \lb1873\footnote{In the \citet{jojor84}
 paper, the primary decay from $^{\rm 4}G^{\rm 0}$ to b$^{\rm 4}G$ 
 produce only \feii\ \lb1870,1873. Nonetheless, data from the Iron
 Project also predict lines at \lb1841 and \lb1845 due to the fine
 structure of that upper level \citep{nahar95, pra01}.
 The current notation for $^{\rm 4}G^{\rm 0}$ is $(u,t)^{\rm 4}G^{\rm 0}$.}.
 A  subsequent cascade to $z^{\rm 4}F^{\rm 0}$
 would produce the 1$\mu$m \feii\ lines, as is illustrated
 in Figure~\ref{levels}. 

 Near-infrared spectroscopy of young stellar objects \citep{hasim88,
 hp89, ham94, kerc94}
 has, in fact, shown most of the near-IR \feii\ emission features
 associated to L$\alpha$ fluorescence. The flux ratios measured 
 from our objects agree with this scenario. 

 \subsubsection{Is Ly$\alpha$ fluorescence also at work in AGNs?}

\begin{figure}
 \plotone{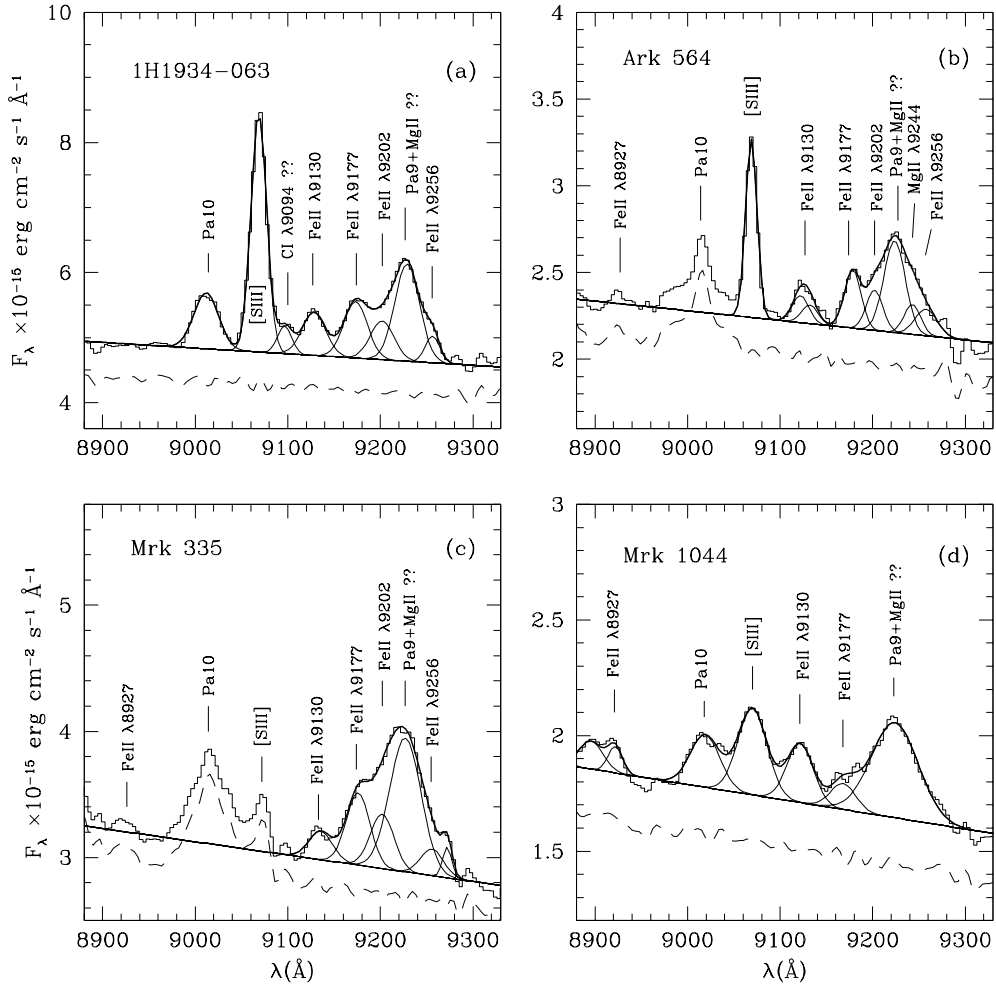}
 \caption{Zoom around the \lb8900$-$\lb9300 region showing the presence 
 of the \feii\ lines at \lb8927, \lb9130, \lb9177 and \lb9202. These lines
 are all primary cascades from L$\alpha$ pumped levels at $\sim$10 eV. 
 \label{feiilyalf}}
 \end{figure}

 In AGNs, a definitive test to confirm the presence of the
 Ly$\alpha$ pumping has been elusive due to both the blending
 of the broad lines and the lack of suitable data in the near-IR region. 
 Recently, \citet[hereafter SP98]{sp98}, 
 using theoretical \feii\ emission-line strengths, derived by 
 considering all of the above four mechanisms, demonstrated that a key 
 feature associated with Ly$\alpha$ pumping is significant \feii\ 
 emission in the wavelength interval 8500$-$9500 \AA. In particular, 
 their models predict strong  \feii\ lines at \lb8490, \lb8927,
 \lb9130, \lb9177 and \lb9203. Figure~\ref{feiilyalf} shows
 an enlarged portion of the \lb8900$-$\lb9300 region for each of 
 the object's spectra, clearly displaying the latter three \feii\ lines 
 and a blip at the expected position of \feii\ \lb8927.
 This result offers the first direct evidence of Ly$\alpha$ 
 pumping in AGNs and will be discussed in detail in \citet{roar01b}.
    
\begin{figure}
 \plotone{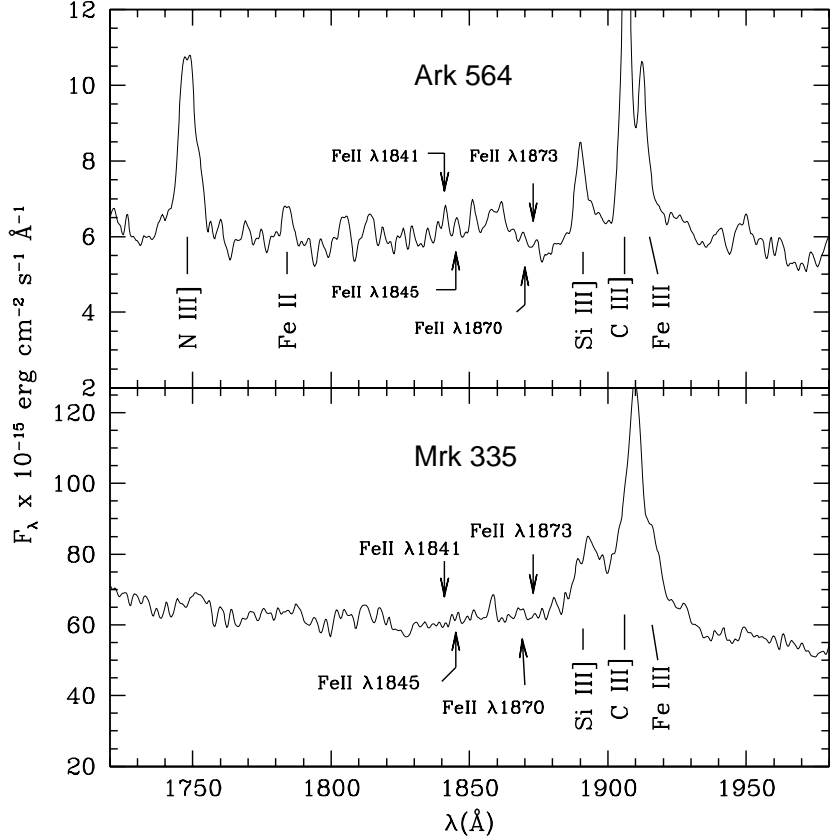}
 \caption{FOS/HST UV spectra for Ark\,564 (top) and Mrk\,335 (bottom) in the
 region around \lb1870. The arrows mark the expected position of the 
 \feii\ lines \lb1841, \lb1845, \lb1870 and \lb1873, resulting from the
 primary decay of the upper $u~^{\rm 4}G^{\rm 0}$ and $t~^{\rm 4}G^{\rm 0}$
 terms, pumped  Ly$\alpha$ fluorescence. The presence of these lines 
 is crucial for determining if this mechanism contributes to the production of 
 the 1$\mu$m \feii\ lines. \label{arkuv}}
 \end{figure}

 If Ly$\alpha$ fluorescence is also responsible for the 
 production of the 1$\mu$m lines, the feed lines \lb1841, \lb1845,
 \lb1870 and \lb1873  
 should be observed (see Figure~\ref{levels}). We have searched 
 the {\it IUE} and {\it HST/FOS} public spectra available for the 
 objects studied here to look for emission lines at these four
 positions. No significant emission was found, as is illustrated
 in Figure~\ref{arkuv}, where the region around \lb1860 in 
 Ark\,564 and Mrk\,335 is shown.
 The spikes detected at the expected positions are almost
 at the noise level. Assuming that they actually correspond to
 \lb1840,1844 and \lb1870,1873, Table~\ref{balance} gives the flux 
 upper limit measured for these features (column 2) and the strength 
 that they should need to have (assuming zero internal reddening) 
 in order to produced the observed 1$\mu$m lines (column 3). 
 The expected values were derived assuming a one-to-one relation
 between the number of photons of the 1$\mu$m and UV \feii\ lines. 
 Column 4 of that same table lists the E(B-V) necessary to reduce
 the expected UV flux to the observed value. It was derived
 using the Galactic extinction law of
 \citet{ccm89}. Since that the amount 
 of reddening should conceal other UV features that
 are otherwise prominent, we discard this effect for explaining the
 weakness of the feed UV lines. Another argument against 
 strong reddening is the presence of a steep blue continuum 
 in most of these objects.  
    
 There is still the possibility that these lines are, in fact,
 emitted but then efficiently destroyed within the BLR. But 
 following the 
 same line of reasoning of RMPH,  other UV features should
 also be affected by such a destruction mechanism. Since
 Ly$\alpha$, \ion{He}{2} $\lambda$1640 and other important
 UV lines are clearly visible, we ruled out 
 that hypothesis. 
 
 We conclude that Ly$\alpha$ fluorescence cannot
 be the only process responsible for the emission of the
 1$\mu$m \feii\ lines. From the observed strength 
 of the UV lines, that mechanism appears to have a minimal
 contribution to the near-IR features.

 \subsubsection{Alternative mechanisms}

 An alternative process to the production of
 the near-IR \feii\ emission is recombination. However,
 in addition to the 1$\mu$m lines,
 a myriad of other transitions should also be present
 in the 0.8 -- 2.4$\mu$m interval, some of them as strong or
 stronger than the ones around 1$\mu$m \citep{kur81}. 
 Their absence clearly rules out this hypothesis.
 
 Continuum fluorescence, first suggested by \citet{waok67},
 has been a popular explanation for the presence of 
 strong \feii\ emission in Seyfert 1 galaxies and quasars.
 In this process, the 5 eV levels of \feii\ are excited by
 absorption of ultraviolet continuum photons in the 
 resonance-line transitions. Nonetheless, this process
 very unlikely contribute to populate the upper b$^{\rm 4}G$ 
 level because no resonant transitions connecting this level 
 to the ground levels have yet been identified.

 There are other two possible excitation mechanisms. The first 
 one takes advantage of the fact that the 1$\mu$m lines
 are, by far, the strongest \feii\ lines observed in the 
 0.8$-$2.4$\mu$m interval. Since they all descend from a common
 upper level ($b~^{4}G$, see Figure~\ref{levels}), their presence
 suggests that some sort of selective mechanism is operating
 to populate that particular level. Collisional pumping from 
 the metastable $a~^{4}G$ level is, at least 
 qualitatively, a viable process to explain the strength of 
 these features in AGNs. The small energy separation between the two
 levels involved ($\sim$3.57 eV), the high density of the
 BLR gas (N$_{\rm e} \leq 10^{9}$ cm$^{-3}$) and the 
 temperature predicted for the \feii\ region by most models 
 ($\sim$8000 K) favors this mechanims. In order to
 be efficient, the population of the level $a~^{4}G$ must
 be significant, i.e. above the Boltzman population.
 This can be done by the decay to $a~^{4}G$ from
 another set of levels such as $^{\rm 4}F^{\rm 0}$.
 A signature of this process is \feii\ emission
 in the 1535$-$1550 \AA\ interval. The \feii\ line at \lb1534 
 (5$p^{\rm 4}F^{\rm 0}-a~^{4}G$) is a particular example of this
 decay route. Unfortunately, its proximity to the strong 
 \ion{C}{4} \lb1548 feature may hamper the detection of that 
 emission. The $a~^{4}G$ levels can also be populated by decays 
 from $y^{\rm 4}H^{\rm 0}$. This route produces lines in 
 the 2430$-$2460 wavelength interval. It is also possible
 that $b^{\rm 4}G$ be collisionally populated from the lower
 $z^{\rm 4}F^{\rm 0}$. That requires high optical depths in 
 the strong optical and UV permitted transitions originating
 from this latter level in order to keep it significantly populated.
 This condition seems plausible because most optical and
 UV \feii\ multiplets are rather strong in AGNs.
  
 The last possible process is to populate $b^{\rm 4}G$ directly 
 by decays from other levels previously pumped by Ly$\alpha$ fluorescence.
 In particular, the upper $^{\rm 4}F^{\rm 0}$ level, after absorption
 of a Ly$\alpha$ photon, may cascade directly to $b^{\rm 4}G$, emitting
 \feii\ lines in the 2700$-$2770 \AA\ interval (\feii\ \lb2707,
 \lb2739, \lb2765, \lb2769). In this case, no collisional excitation
 of $b^{\rm 4}G$ is required. An inspection of the {\it IUE/HST} 
 spectra of Ark\,564, Mrk\,335 and Mrk\,1044 show evidence of
 the above lines at the expected position, suggesting that this
 process may also contribute to the 1$\mu$m \feii\ 
 line intensities.

 \subsubsection{Conclusions} 
 
 In summary, clear evidence of Ly$\alpha$ pumping of \feii\ 
 is found from the observation of the primary cascade lines in
 the 8900$-$9300 \AA\ interval, formed by the decay of the upper 5p 
 levels to the $e^{\rm 4}D$ and $e^{\rm 6}D$ levels. This result
 confirms that fluorescence excitation by Ly$\alpha$ has an
 active role in the \feii\ spectra of AGNs. Nonetheless,
 contrary to what was previously suggested, this same process seems 
 to be insufficient to explain the 
 strength of the 1$\mu$m lines, the strongest of all \feii\ lines 
 in the 8000$-$24000 \AA\ region. This conclusion is drawn from the 
 weakness of the primary cascade lines that feed the upper term from which
 the 1$\mu$m \feii\ emission originates. Our observations point 
 to a combination of collisional excitation and decays from 
 levels populated by Ly$\alpha$ photons to 
 produce the \feii\ spectrum in the near-IR. 
  
 \subsection{Relative Intensities of the 1$\mu$m \feii\ Lines}

 One advantage of the 1$\mu$m \feii\ lines over the
 corresponding emission in the optical region is that 
 most of the individual lines are clearly separated in
 wavelength, allowing their measurement without the need
 of constructing an \feii\ template to
 quantify them. Although in most cases emission from 
 \ion{H}{1}, \ion{He}{1} and \ion{He}{2} is near the \feii\ 
 lines, it is still possible to measure, at a good 
 confidence level, the flux of the individual components. 

 In all cases our spectra show that $\lambda$9997 
 is the strongest \feii\ transition in the near-IR. The 
 relative fluxes of other \feii\ lines compared 
 to that of $\lambda$9997 are listed in Table~\ref{feiirat}. 
 We have also included for comparison, data taken 
 from RMPH for I\,Zw\,1 (column 6), from \citet{rudy91} 
 for LkH$\alpha$\,101 (column 7), and from \citet{kerc94} for
 MWC 249 and R\,Mon (columns 8 and 9 respectively). The
 last three objects are Galactic sources, well known for 
 their strong \feii\ emission lines in the near-IR,
 which is believe to arise in a gas with physical
 conditions similar to those found in AGNs.

 The 1$\mu$m \feii\ ratios vary from 
 object to object, as is seen in  Table~\ref{feiirat}. Comparing 
 the values derived for our sample with those of I\,Zw\,1 
 (considered as the prototype NLS1), we found that, within errors, 
 1H\,1934-063 is the most I\,Zw\,1-like. On the other hand, Mrk\,1044
 deviates most significantly in regard to the \feii\
 emission. The situation is
 also highly discrepant when we compare the NLS1 ratios
 to those of Galactic objects. No agreement between these two
 groups can be seen. The ratios among
 the Galactic sources look rather constant, in contrast to
 the large scatter showed by the NLS1 data. \citet{rudy91}
 had already noted the similarity of line ratios between
 Lk\,H$\alpha$ 101 and $\eta$ Car, another Galactic source,
 and argued that this may suggest that the same process gives
 rise to the lines in both. They proposed Ly$\alpha$ fluorescence 
 as the dominant mechanism for the production of \feii\ 
 but did not discard collisional excitation. 

\begin{figure}
 \plotone{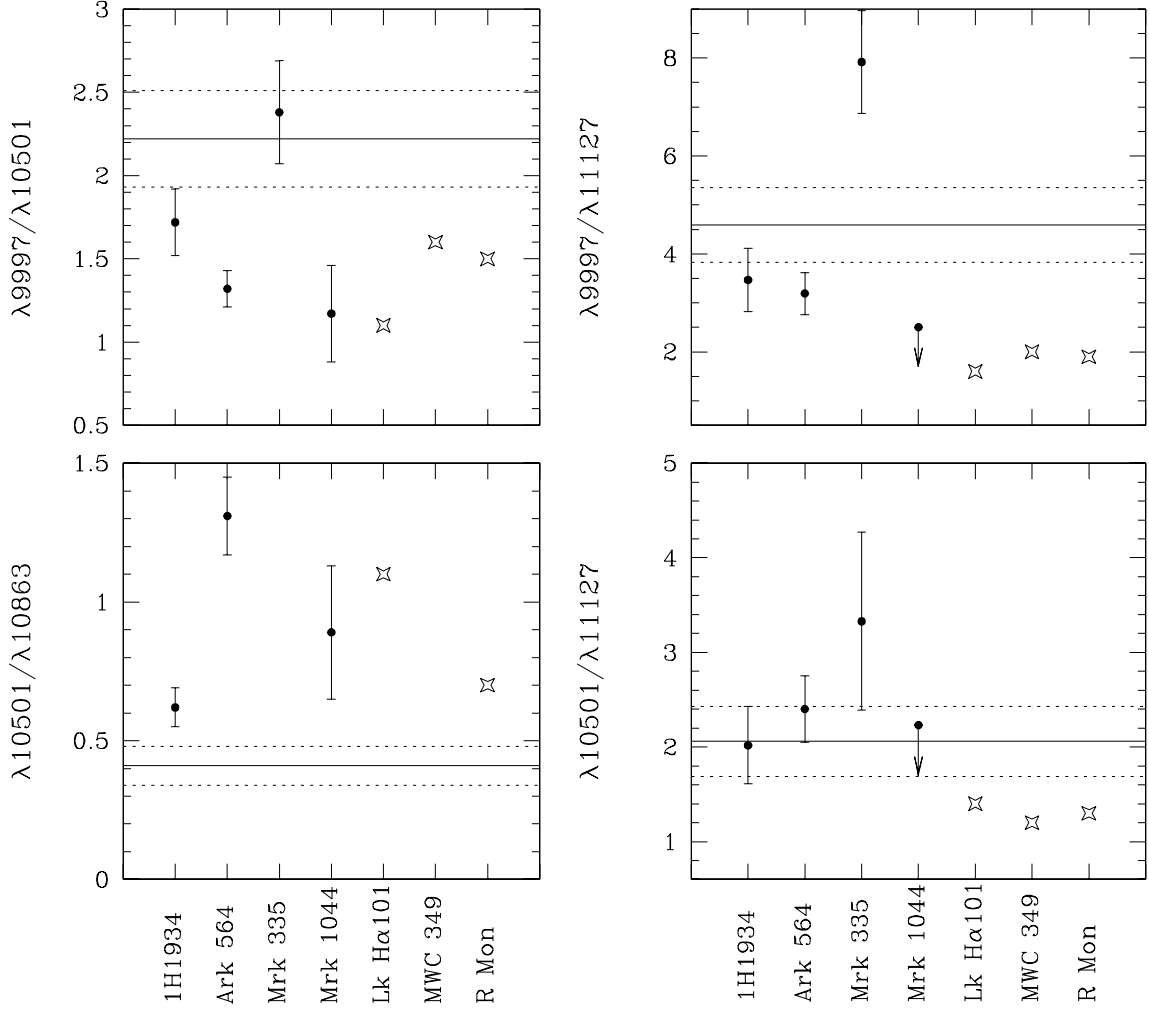}
 \caption{Ratios between the most important \feii\ lines
 observed in the galaxies and the Galactic sources 
 Lk\,H$\alpha$101, MWC\,349, and R Mon. The solid line is the value reported
 by RMPH for I\,Zw\,1 and the dotted line corresponds to the uncertainty.
 The starred points correspond to the Galatic sources. 
 \label{ratiosfeii}}
 \end{figure}

 It may be argued that the large scatter in \feii\ line 
 ratios is due to reddening, since none of the NLS1 
 spectra  have been corrected for internal extinction
 (the Galactic source line ratios all have been corrected). 
 However, the near-IR region is less affected by 
 reddening than the UV and optical regions  and the wavelength
 interval where the 1$\mu$m \feii\ features are located 
 is relatively small. Thus, one would expect that the 
 reddening is not important here. In addition, all the 
 objects have strong emission lines in the UV, 
 indicating that the reddening
 is small. In order to completely avoid any reddening effect,
 we have calculated line ratios that are very 
 close in wavelength and therefore insensitive to reddening.
 Such is the case for \feii\ \lb9997/\lb10171, \feii\ \lb11127/\lb10863 and 
 \feii\ \lb10501/\lb10863. The results, also listed in 
 Table~\ref{feiirat}, follow the same trend, in the 
 sense that the line ratios vary from galaxy to galaxy 
 and also differ from the line ratios derived for the Galactic 
 sources. This variation is more clearly seen in 
 Figure~\ref{ratiosfeii}, where the line ratios are plotted 
 for the objects listed in Table~\ref{feiirat}. The large scatter 
 of the NLS1 data and the constancy among Galactic sources
 is evident.  

\begin{figure}
 \plotone{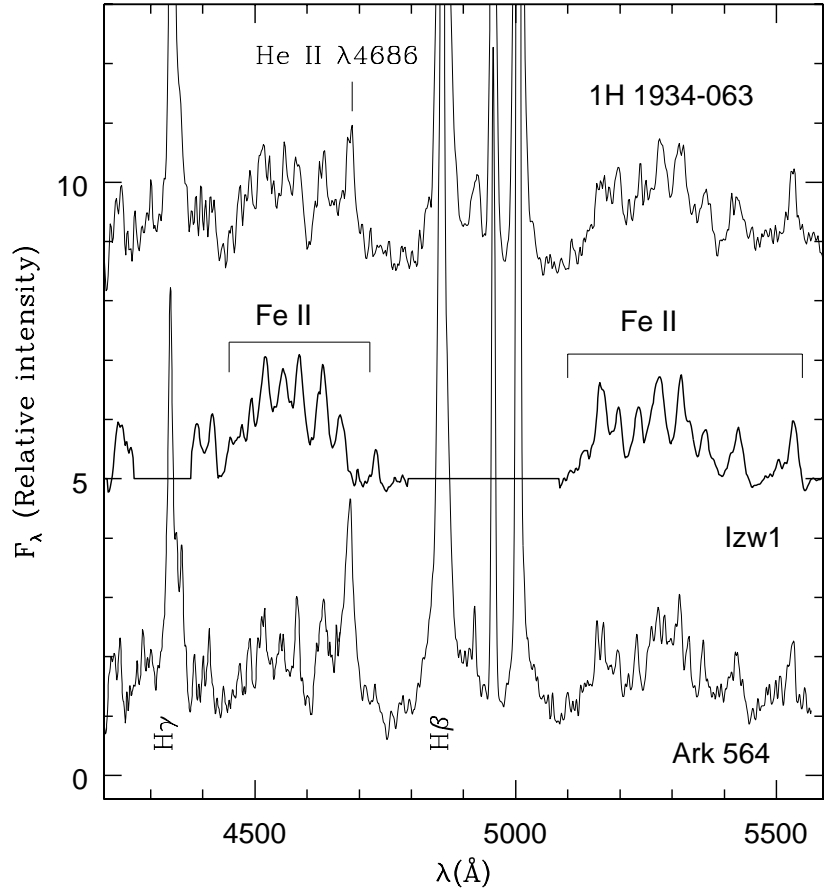}
 \caption{Comparison of the optical \feii\ spectra of 1H\,1934-063
 (upper), I\,Zw\,1 (middle) and Ark\,564 (lower). Except for the
 width of the individual lines (narrower in Ark\,564), the three
 spectra look similar. Note that the relative intensities of 
 the \feii\ lines seem to be constant in the three objects.
 The H$\beta$ and H$\gamma$ lines of I\,Zw\,1 were removed
 for clarity.  
 \label{optfeii}}
 \end{figure}

 The above results strongly contrast to what is observed
 in the optical region, where the relative intensities of the
 \feii\ lines in AGNs seems rather similar from object to 
 object. Apparently, the only difference between the optical \feii\
 features of two given objects is the width of the individual 
 lines (unless a scale factor). This result was clearly illustrated 
 in \citet{phi78} by comparing the \feii\ emission in the region 
 4400$-$5300 \AA\ of several Seyfert 1 galaxies. The \feii\ 
 emission in that same interval for I\,Zw\,1,
 Ark\,564 and 1H\,1934-063 is shown in Figure~\ref{optfeii}.
 Note how similar are the \feii\ multiplets
 in these objects. For that reason, usually I\,Zw\,1, which has
 one of the narrowest permitted lines in Seyfert 1s, is used as a 
 template to determine the strength of the \feii\ emission 
 \citep{bogre91, roar00}.

\begin{figure}
 \plotone{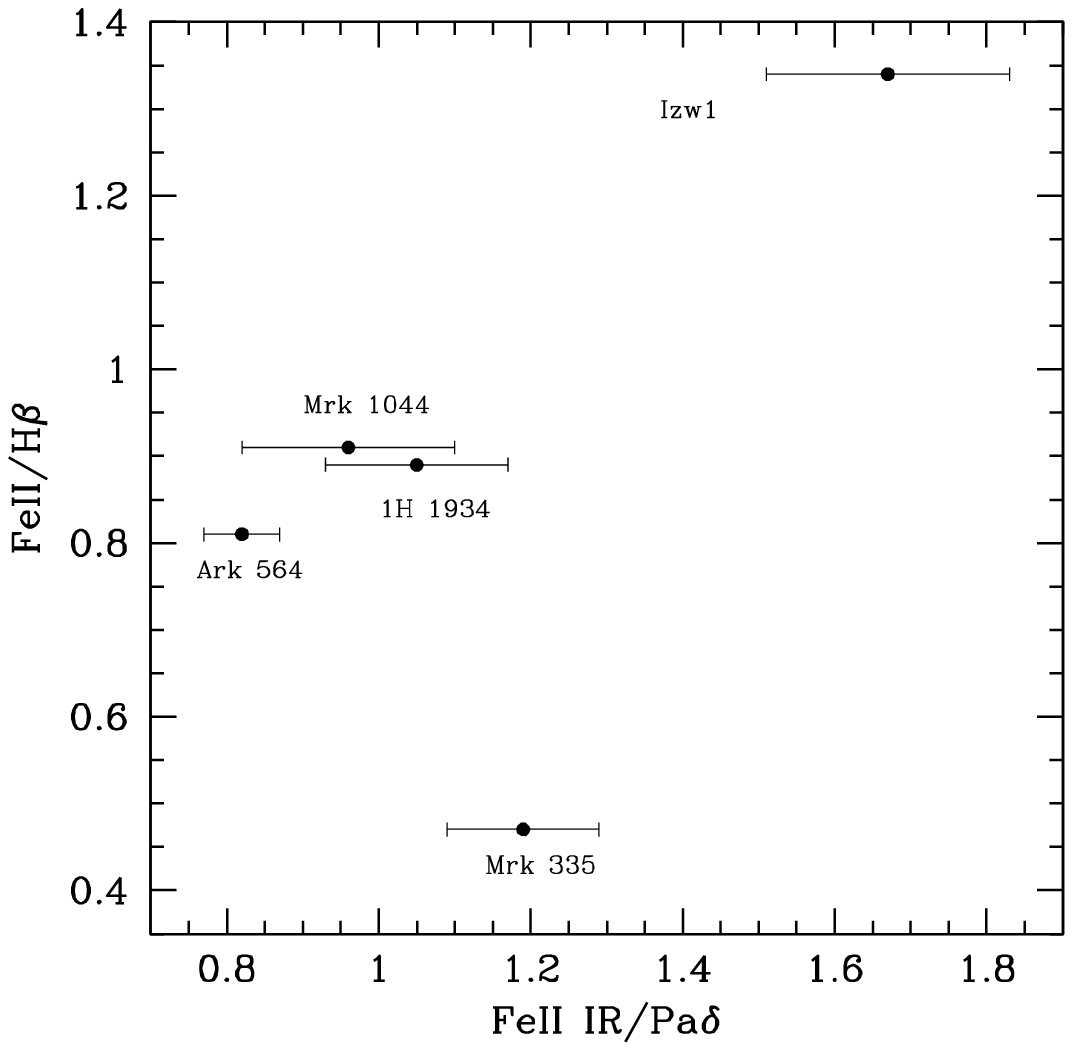}
 \caption{Optical \feii\ emission (\feii \lb4750/\hbeta) versus the
 sum of the flux of \feii\ \lb9997, \lb10171 and \lb10501 divided by Pa$\delta$
 (\feii\ IR/Pa$\delta$). Values of \feii \lb4750/\hbeta\ are listed in Table~\ref{feiirat}.
 Mrk 335 clearly departs from an apparent correlation 
 between these two quantities.  \label{feiicorr}}
 \end{figure}

 The question that arises now is if there 
 is a correlation between the strength of the \feii\ in 
 the optical and near-IR. Figure~\ref{feiicorr} shows 
 FeII/\hbeta\ versus \feii\ IR/Pa\,$\delta$. The former is 
 the flux of the \feii\ emission centered in $\lambda$4570, 
 which scales with the total optical \feii, normalized to 
 \hbeta, while the latter is the sum of the \feii\ lines
 \lb9997, \lb10171 and \lb10501, normalized 
 to the Pa$\delta$ flux. Except for Mrk\,335 (the galaxy with the
 smallest \feii/\hbeta), \feii\ \lb9997 seems to scale
 with the corresponding optical \feii. That correlation is
 very important, since most of the optical \feii\ lines 
 in the wavelength interval 4400$-$5300  arise from 
 the same upper two upper levels $z^{\rm 4}F^{\rm 0}$ and 
 $z^{\rm 4}D^{\rm 0}$. Although no definitive conclusions
 can be drawn due to the reduced number of points, 
 Figure~\ref{feiicorr} clearly suggest that the optical
 \feii\ emission is directly linked to the 1$\mu$m \feii\
 lines. In this scenario, decays from $b^{\rm 4}G^{\rm 0}$ largely 
 contribute to the population of those two levels. Although
 a detailed balance between the energy carried by the near-IR
 and optical \feii\ lines is far from the scope of this paper,
 this result is important for current \feii\ modeling.     
 
 Factors such as different optical depth effects and, possibly,
 different sensitivities to physical parameters between the
 near-IR and optical \feii\ lines can be invoked
 to explain why the line ratios in the former region present
 more scatter than those of the latter. Additional observations
 in a larger sample of objects and a more accurate modeling 
 of the 1$\mu$m \feii\ lines are needed to explain this trend. 
 It is also interesting 
 to understand why Mrk\,335 deviates so much from the other 
 NLS1s. While its optical \feii\ is rather low for a typical 
 NLS1s, its near-IR \feii\ ranks it as a strong 
 emitter. 

 \section{Kinematics of the BRL region} \label{kin} 

 The quality and spectral resolution of the SpeX data
 allow us to study and compare, for the first time, the 
 form and width of the different near-IR permitted emission 
 line profiles of AGNs, and in particular of NLS1s. This
 is important for at least two reasons. First,
 it has been argued (from both theoretical models and 
 observational grounds) that low-ionization lines such as 
 \feii, \ion{O}{1}, \ion{Mg}{2}, \ion{Ca}{2} and Balmer
 lines are formed in a distinct, separate zone (most
 probably the outer part of the BLR) from the place 
 where most high-ionization lines (i.e Ly$\alpha$, 
 \ion{He}{2}, \ion{Ca}{4}) are produced \citep{csdjp86, rms97}. 
 Differences in line widths between low 
 and high ionization lines are usually interpreted as 
 evidence of this stratification, provided the velocity 
 field does not differ strongly from gravitationally
 bound Keplerian or virial motion, as seems to be 
 the case for most AGNs \citep{pw99}. 
 Second, the presence of asymmetries in the line 
 profiles  helps to distinguish between radial and 
 rotational motions, setting constrains on the
 velocity field of the BLR. 

 Although the visible and UV spectral regions in AGNs 
 are full of permitted emission lines, usually they are 
 severely blended, affecting the placement of 
 the continuum level and the separation of the individual
 profiles. Such is the case with the \feii\ 
 lines, which form a pseudo-continuum even in NLS1s
 because of their close proximity. The near-IR region is
 promising in this respect (if the resolution is high enough).
 Lines such as \ion{Ca}{2} 
 \lb8664, \feii\ \lb11127, \ion{O}{1} \lb11287 and 
 Pa$\beta$ are completely isolated or mildly blended, 
 making their separation and the analysis of their 
 emission profiles easier.

 For the above reasons we have chosen \feii\ \lb11127, 
 \ion{O}{1} \lb11287, \ion{O}{1} \lb8446 and \ion{Ca}{2} 
 \lb8664 as representative of low-ionization BLR lines, and 
 Pa$\beta$ as an example of a BLR recombination line. 
 By comparing their emission line profiles, we expect to
 get clues about the structure and kinematics of  
 the emitting region. In addition,  
 we have used the [\ion{S}{3}] \lb9531 profile to
 represent the velocity field of the Narrow Line
 Region (NLR).

\begin{figure}
 \plotone{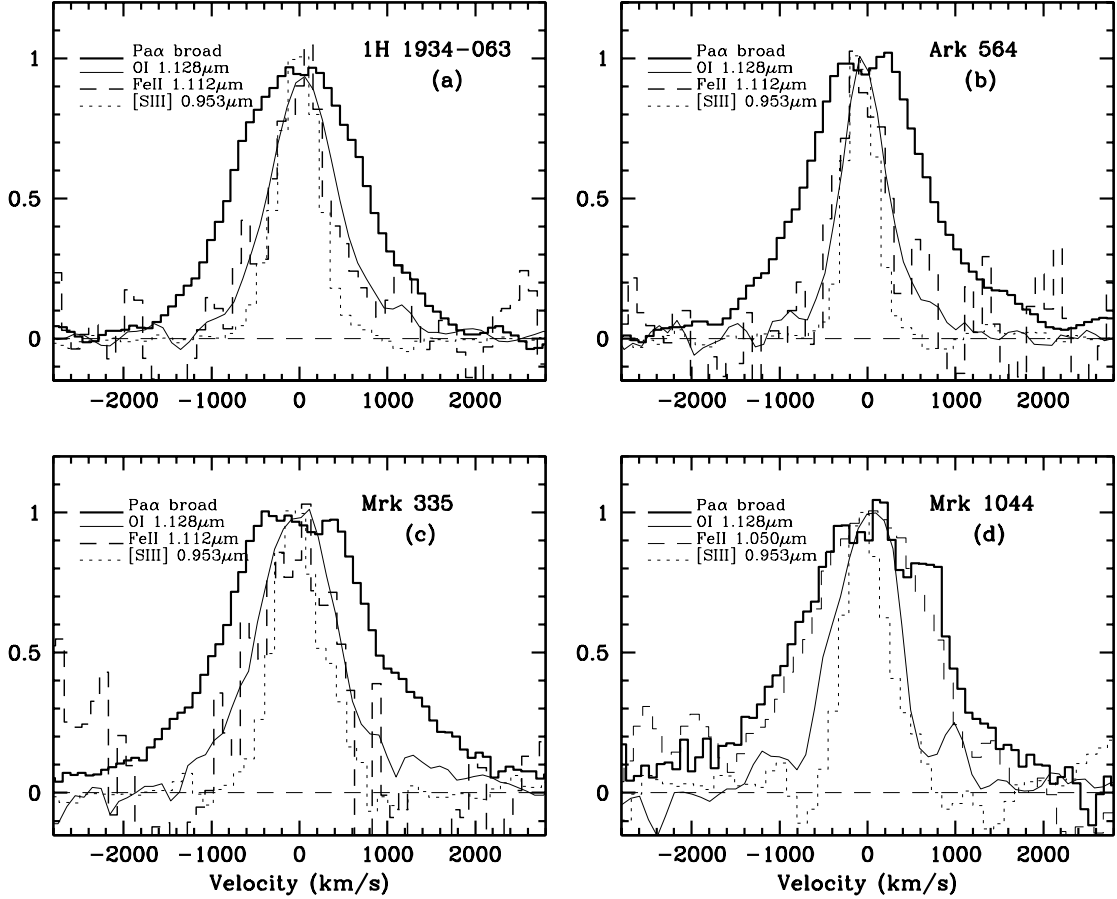}
 \caption{Comparison of the line profiles of Pa$\beta$ (thick histogram), 
 \ion{O}{1} \lb11287 (solid line), \feii\ \lb11127 (dashed histogram) and
 [\ion{S}{3}] \lb9531 (dotted histogram) in velocity space, showing clearly 
 that Pa$\beta$ is significantly wider than the lines emitted
 in the partially ionized zone of the the BLR. This suggests that 
 low-ionization lines such as \feii\ and \ion{O}{1} form in
 the outermost part of the BLR.
   \label{profiles}}
 \end{figure}

 The observed profiles of Pa$\beta$, \feii\ \lb11127,
 \ion{O}{1} \lb11286 and [\ion{S}{3}] \lb9531 are shown 
 in velocity space in Figure~\ref{profiles}. 
 In order to easily compare them, the peak 
 intensity of each line has been normalized to unity. 
 In all cases, Pa$\beta$ has been stripped of the 
 narrow component by subtracting the contribution 
 from the NLR according to the procedure explained 
 in \S \ref{observa}. Since the \feii\ and 
 the \ion{O}{1} emission are features restricted to 
 the BLR, there is no need to determine such a 
 contribution in these two lines.

 The results are surprising. Except for Mrk\,1044,
 \feii\ and \ion{O}{1} share almost the same velocity 
 field and display very similar line profiles. However, 
 they are significantly narrower than Pa$\beta$. In
 fact, they are just slightly broader than [\ion{S}{3}]
 \lb9531. Assuming that this latter line is formed
 in the inner portion of the NLR (its critical density,
 $N_{c}$, is $\sim 10^{7.5}$ cm$^{-3}$), 
 the small width of \feii\ and \ion{O}{1} indicates
 that we are probably looking at the outer boundaries
 of the BLR. The result obtained for Mrk\,1044 can
 be explained if we recall that in 
 Figure~\ref{profiles} the \feii\ profile plotted 
 corresponds to that of \lb10501 instead of \lb11127, because
 this latter line is severely affected by atmospheric
 absorption (see \S \ref{mrk1044}). 
 \feii\ \lb10501 is, in fact, a blend of two \feii\ lines 
 located at \lb10491 and \lb10501. Observations in
 stars \citep{rudy91} show that the latter line is about 5 times 
 stronger than the former (in which case the effect of \lb10491
 on the blend width would be negligible). A lower ratio may cause 
 \lb10501 to appear broader, as it seems to be the case 
 for Mrk\,1044 as well as for Ark\,564 
 (see Table~\ref{tableflux}).

 Assuming that the clouds emitting the broad lines 
 are gravitationally bound to the central mass 
 concentration and that the velocity field is Keplerian, 
 Figure~\ref{profiles} clearly supports the hypothesis 
 that \feii\ and \ion{O}{1} are emitted in the outer part 
 of the BLR while most of the Pa$\beta$ flux is formed 
 in the intermediate portion of this region. 
 This result is compatible with the physical conditions 
 needed to form permitted \feii\ and \ion{O}{1}, i.e. neutral 
 gas strongly shielded from the incident ionizing 
 radiation of the central source, conditions which can
 only be met deeper within the neutral zone of the
 outermost part of the BLR. 

 The above results seem to contradict that of \citet{morward89}, 
 who compared the H$\alpha$ and \ion{O}{1} \lb8446 line profiles 
 in a sample of 12 objects. They found no evidence for 
 differences between the profiles of these two lines. 
 Nonetheless, recall that Pa$\beta$ is less affected than 
 H$\alpha$ by radiation transport effects. When crossing the BLR,
 H$\alpha$ photons produced in inner higher velocity clouds 
 are more likely to be absorbed than Pa$\beta$ photons. Thus, 
 the H$\alpha$ line probes the outer low velocity clouds, while  
 the Pa$\beta$ profile shows the contribution from inner high
 velocity clouds. For the above reason, we expect that
 Pa$\alpha$ be broader than H$\alpha$ as well as broader
 than low ionization lines such as \ion{O}{1} and \feii.
 The most important point here is that Figure~\ref{profiles}
 shows that low-ionization BLR lines are emitted almost
 at the border between the BLR and NLR.

 Another interesting result that can be drawn from
 Figure~\ref{profiles} is the difference in the
 form of the profiles from object to object: in 
 1H\,1934-063 they are typically Gaussian while in Ark\,564 
 they are mostly Lorentzian, with Mrk\,335 and Mrk\,1044
 being intermediate cases. This is valid not only for Pa$\alpha$
 but also for \feii\ and \ion{O}{1}. Since the spectra were 
 taken at the same resolution, with the same telescope and setup, this
 clearly indicates differences in the velocity fields
 of the objects. 

\begin{figure}
 \plotone{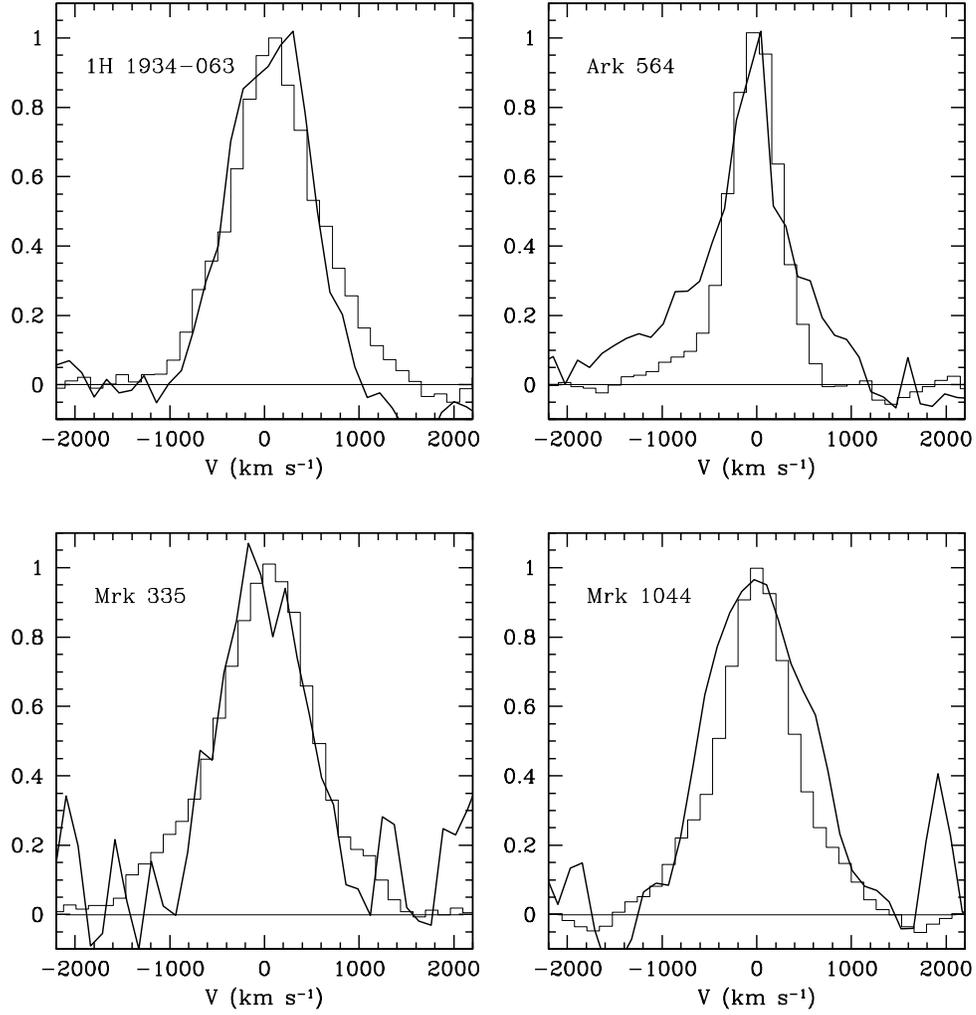}
 \caption{Comparison of the emission line profiles of \ion{Ca}{2}
 \lb8662 (thick line) and \ion{O}{1} \lb8446 (histogram), 
 in velocity space. \label{caii}}
 \end{figure}

 It is also possible to probe weakly ionized material within
 the BLR by means of \ion{Ca}{2} \lb8664, one of the Ca
 near-IR triplet lines that is perfectly isolated and has 
 good S/N in our data. The observed \ion{Ca}{2} \lb8664 
 is compared to that of \ion{O}{1} \lb8446 in Figure~\ref{caii}.
 Clearly, these two lines have very similar form and width, 
 indicating that they share the overall 
 kinematics of the region where they are
 emitted. This was already pointed out by \citet{pe88} in 
 his study of the \ion{Ca}{2} and \ion{O}{1} lines in
 14 out of 40 AGNs for which simultaneous emission of 
 these two ions was found. Here, our results not only
 agree with those of \citet{pe88}, but also extend
 that analysis to the \feii\ and show that those 
 three lines should have similar kinematics.
 Due to the high complexity the \feii\ atom, its spectrum 
 is far from being understood. Information derived for
 \ion{O}{1} or \ion{Ca}{2} (i.e. column density and ionization
 parameters) provide valuable constraints for modelling the \feii\
 atom, given that these species seem to form from
 the same ensemble of clouds.

 The importance of the \ion{Ca}{2} 
 emission as a valuable diagnostic of the the BLR is
 stressed by \citet{ferper89}. Using photoionization
 models that reproduce well both the high and low ionization
 line intensities, they found that clouds with extremely
 high column densities (N$_{\rm e} \sim$ 10$^{25}$ cm$^{-2}$)
 are needed to emit \ion{Ca}{2}. The large N$_{\rm e}$ suggests
 that the clouds might be a wind or corona above a star or
 accretion disk. It is well known that a wind scenario 
 usually produces asymmetries in the observed profiles.
 The \ion{Ca}{2} line profiles in Ark\,564 marginally
 show evidences of assymetries towards the blue (Figure~\ref{caii}, 
 upper right panel). High-resolution spectroscopy around
 this region is highly desirable to find additional clues 
 about the kinematics of the outer portion of the BLR.
 Our results stress the kinematic connection 
 between the low ionization species and confirm previous
 claims that the understanding of the mechanisms leading to 
 the formation of the \feii\ and \ion{Ca}{2} lines is crucial 
 for solving the puzzle associated with the geometry and 
 physical conditions of the outer parts of the BLR.  
   
 In summary, the comparison of the near-IR broad emission 
 line profiles definitively shows the presence of a stratification
 within the BLR. Low-ionization lines such as \feii, \ion{O}{1}
 and \ion{Ca}{2} are emitted in the outer edge of the BLR. 
 Pa$\beta$ emission, on the other hand, maps the velocity field 
 of the inner portions of the BLR. 
  
 \section{Conclusions}

 We have analyzed four NLS1 spectra in the wavelength range
 0.8 -- 2.4$\mu$m in order to study the \feii\ emission and
 the kinematics of the BLR. The results show that the \feii\
 emission in the 8500$-$9300 \AA\ interval and around 
 1$\mu$m are common features of this class of AGNs. The
 \feii\ lines in the former interval are, for the first time, 
 resolved in this type of object. Since they are primary cascades of the
 upper 5p levels pumped by Ly$\alpha$ fluorescence, our
 data offer clear observational evidence for the presence of 
 this excitation mechanism in AGNs. The \feii\ lines located
 at \lb9997, \lb10501, \lb10863 and \lb11127, which share
 the same upper and lower levels ($b^{\rm 4}G$ and $z^{\rm 4}F^{\rm 0}$,
 respectively), are very prominent and the strongest \feii\ lines 
 in the four spectra. Although Ly$\alpha$ fluorescence
 indirectly contributes to their production, that
 mechanism cannot explain their full strength because of the 
 weakness of the UV \feii\ lines that feed the upper $b^{\rm 4}G$
 level. Collisional ionization, mainly from the $a^{\rm 4}G$ and 
 $z^{\rm 4}F^{\rm 0}$ levels, is suggested as an additional
 mechanism that plays an active role in exciting the 1$\mu$m
 \feii\ lines.
    
 The near-IR \feii\ line ratios (i.e. \lb9997/\lb10501 and 
 \lb10501/\lb10863) show a large scatter from 
 object to object, contrary to what is observed in the 
 optical region. Nonetheless, the apparent good correlation 
 found between the \feii\ emission of these two wavelength 
 intervals supports a common origin. Since most of
 the optical \feii\ multiplets originate in transitions
 from $z^{\rm 4}F^{\rm 0}$, that level would be mainly
 populated by the decay from $b^{\rm 4}G$.  

 The emission line profiles of \feii,
 \ion{O}{1} and \ion{Ca}{2} are very similar in form and
 width but significantly narrower than that of Pa$\beta$. 
 This strongly suggests that the \ion{Ca}{2}, \feii\ and
 \ion{O}{1} regions are co-spatial and kinematically linked.
 Moreover, because their profiles are just slightly broader
 than that of [\ion{S}{3}] \lb9531, low ionization BLR lines
 are most likely formed in the outermost portion of
 the BLR. This is compatible with the physical conditions
 required to form them: cool gas with high column 
 densities that provides a shield from the incident 
 ultraviolet continuum of the central source and a location
 deep, within the neutral zone of the BLR. The larger
 width of the Paschen lines indicates that they are
 suitable for mapping the gas located at intermediate positions
 in the BLR.  
    
 Given the high complexity of the \feii\ atom and its ubiquitous 
 emission in AGNs, only detailed model calculations that take into 
 account all possible excitation mechanisms and transitions among
 the levels can shed some light on the dominant processes involved
 in the excitation of this atom. The data presented here 
 provide valuable constraints for understanding this complex emission.

 \acknowledgments

We thank the referee, Dr. Fred Hamann, for his useful and 
thoughtful comments that helped to improve this manuscript. 
We also wish to thank the IRTF staff, support scientist Bobby Bus, 
and telescope operator Dave Griep for contributing to a productive 
observing run. Mike Cushing provided patient assistance with the 
xspextool software. This research has been supported by the 
Funda\c c\~ao de Amparo a Pesquisa do Estado de S\~ao Paulo $-$
FAPESP, under contract 00/01020-5 and PRONEX grants 662175/1996-4 
and 7697100300. We acknowledge the use the Atomic
Line List v2.04 (http://www.pa.uky.edu/~peter/atomic/).

 \clearpage

 \clearpage

 \clearpage

 \clearpage

 \clearpage

 \clearpage

 \clearpage

 \clearpage

 \clearpage

 \clearpage

 \clearpage

 \clearpage

 \clearpage

 \clearpage

 \begin{deluxetable}{lcccc}
 \tablecaption{Basic information of the NLS1 galaxies \label{tabledata}}
 \tablewidth{0pt}
 \tablehead{
 \colhead{Galaxy} & \colhead{z} & \colhead{M$_{v}$\tablenotemark{a}} &

 \colhead{A$_{v}$\tablenotemark{b}} & {Morphology}
 }
 \startdata
 1H\,1934-063 & 0.01059 &  -19.04   & 0.972 & Sb  \\
 Ark\,564     & 0.02468 &  -20.42   & 0.198 & SB  \\
 Mrk\,335     & 0.02578 &  -21.32   & 0.118 & S0/a \\
 Mrk\,1044    & 0.01645 &  -18.84   & 0.113 & SB0 \\
 \enddata
 \tablenotetext{a}{A value of H$_{\rm o}$ = 75 \kms Mpc$^{-1}$ was assumed}
 \tablenotetext{b}{Galactic extinction}

 \end{deluxetable}

 \clearpage

 \begin{deluxetable}{lcccccccc}
 \tabletypesize{\scriptsize}
 \tablecaption{Observed emission line fluxes and FWHM for the 
 permitted lines\tablenotemark{a}. \label{tableflux}}
 \tablewidth{0pt}
 \tablehead{
 \colhead{} & \multicolumn{2}{c}{1H\,1934-063} & \multicolumn{2}{c}{Ark\,564} &
 \multicolumn{2}{c}{Mrk\,335}  & \multicolumn{2}{c}{Mrk\,1044} \\
 \colhead{Line} & \colhead{Flux} & \colhead{FWHM} & \colhead{Flux} &
\colhead{FWHM} & \colhead{Flux} & \colhead{FWHM} & \colhead{Flux} & \colhead{FWHM}
 }
 \startdata
 \ion{Fe}{2} \lb8927  & \nodata  & \nodata  & 0.18$\pm$0.05 &580  & 0.34$\pm$0.13 & 700 & 0.22\tablenotemark{2} & \nodata \\
 \ion{Fe}{2} \lb9127  & 2.13$\pm$0.41 & 950 & 0.59$\pm$0.11 &600  &0.72$\pm$0.23  & 900 & 0.91$\pm$0.29 & 1050  \\
 \ion{Fe}{2} \lb9177  & 2.77$\pm$0.41 & 950 & 0.75$\pm$0.15 &600  & 1.83$\pm$0.23 & 900 & 0.38$\pm$0.25 & 1050 \\
 \ion{Fe}{2} \lb9202  & 1.90$\pm$0.41 & 950 & 0.52$\pm$0.15 &600  & 1.38$\pm$0.23 & 900 & \nodata & \nodata \\
 Pa9 \lb9229          & 5.00$\pm$0.42 & 950 & 1.91$\pm$0.20 &820  & 4.52$\pm$0.31 & 1250 & 2.60$\pm$0.47 & 1900  \\
 \ion{Fe}{2} \lb9256  & 0.80$\pm$0.25 & 500 & 0.46$\pm$0.20 &800  & 0.66$\pm$0.21 & 900 & \nodata & \nodata \\
 \ion{Fe}{2} \lb9956  & 1.24$\pm$0.28 & 1000 &0.54$\pm$0.07 &750  & \nodata & \nodata    & 0.63$\pm$0.07 & 1000  \\
 \ion{Fe}{2} \lb9997  & 4.52$\pm$0.28 & 1000& 1.72$\pm$0.07 &750  & 4.36$\pm$0.12 & 1600 & 1.36$\pm$0.07 & 1000  \\
 \ion{He}{2} \lb10124 & 3.44$\pm$0.30 & 1100& 1.72$\pm$0.13 &1650 & 3.62$\pm$0.14 & 1850 & 1.06$\pm$0.14 & 1800  \\
 \ion{Fe}{2} \lb10171 & 0.85$\pm$0.28 & 1000& 0.38$\pm$0.07 &750  & 0.77$\pm$0.12 & 1600 & 0.22$\pm$0.07 & 1000  \\
 \ion{Fe}{2} \lb10501 & 2.62$\pm$0.25 & 1000& 1.30$\pm$0.09 &1100 & 1.83$\pm$0.23 & 1600 & 1.16$\pm$0.28 & 1620   \\
 \ion{Fe}{2} \lb10863 & 4.24$\pm$0.25 & 950 & 0.99$\pm$0.08 &700  & \nodata &\nodata     & 1.31$\pm$0.17 & 900  \\
 \ion{Fe}{2} \lb11127 & 1.30$\pm$0.23 & 850 & 0.54$\pm$0.07 &600  & 0.55$\pm$0.14 & 1100 & 0.52\tablenotemark{3}      \\
 \ion{O}{1} \lb11287 &  6.91$\pm$0.24 & 850 & 2.70$\pm$0.07 &600  & 3.87$\pm$0.17 & 1100 & 1.52$\pm$0.17 & 900  \\
 Pa$\alpha$ \lb18761\tablenotemark{3} & 25.43 & 970 & 13.46 &600  & 20.76 & 1200& 4.21          & 1200  \\
 Br$\gamma$ \lb21655 & 2.24$\pm$0.14  & 850 & 1.05$\pm$0.06 &600  & 2.67$\pm$0.31 & 1150 & 0.78$\pm$0.91 & 1000   \\
 \enddata
 \tablenotetext{a}{Fluxes in units of 10$^{-14}$ erg cm$^{-2}$ s$^{-1}$ and FWHM
in \kms}
 \tablenotetext{1}{lower limit}
 \tablenotetext{2}{upper limit}
 \tablenotetext{3}{Flux measurement affected by atmospheric absorption, so it represents a
 lower limit}

\end{deluxetable}

\clearpage

 \begin{deluxetable}{lcccccccc}
\tabletypesize{\scriptsize}
 \tablecaption{Observed emission line fluxes and FWHM for the 
 permitted lines with broad and narrow components\tablenotemark{a}. \label{tablefluxbn}}
 \tablewidth{0pt}
 \tablehead{
 \colhead{} & \multicolumn{2}{c}{1H\,1934-063} & \multicolumn{2}{c}{Ark\,564} &
 \multicolumn{2}{c}{Mrk\,335}  & \multicolumn{2}{c}{Mrk\,1044} \\
 \colhead{Line} & \colhead{Flux} & \colhead{FWHM} & \colhead{Flux} &
\colhead{FWHM} & \colhead{Flux} & \colhead{FWHM} & \colhead{Flux} & \colhead{FWHM}
 }
\startdata
 Pa$\delta_{\rm n}$ \lb10049    & \nodata    & \nodata & 1.00$\pm$0.03 & 320  & \nodata   & \nodata   & \nodata   & \nodata \\
 Pa$\delta_{\rm b}$ \lb10049    & 7.58$\pm$0.27 & 1000 & 3.16$\pm$0.10 & 1650 & 5.86$\pm$0.09& 1100 & 2.85$\pm$0.09  & 1250 \\
 \ion{He}{1}$_{\rm n}$ \lb10830 & 7.53$\pm$0.20 & 410  & 4.90$\pm$0.08 & 370  & 5.49$\pm$0.14  & 550 &  \nodata   & \nodata \\
 \ion{He}{1}$_{\rm b}$ \lb10830 & 19.6$\pm$0.33 & 1230 & 5.46$\pm$0.41 & 2400 & 24.10$\pm$0.44 & 2020 & 8.35$\pm$0.19 & 1000 \\
 Pa$\gamma_{\rm n}$ \lb10937   & 4.29$\pm$0.25 & 570  & 1.84$\pm$0.07 & 320  & \nodata   & \nodata  &  \nodata   & \nodata \\
 Pa$\gamma_{\rm b}$ \lb10937   & 9.22$\pm$0.41 & 1600 & 4.00$\pm$0.29 & 1700 & 11.60$\pm$0.24 & 1500 & 3.92$\pm$0.28 & 1500 \\
 Pa$\beta_{\rm n}$ \lb12820   & 8.36$\pm$0.22 & 600  & 3.95$\pm$0.08 & 400  & 4.29$\pm$0.13  & 600  & 2.22$\pm$0.10 & 600 \\
 Pa$\beta_{\rm b}$ \lb12820   & 9.64$\pm$0.39 & 1800 & 5.90$\pm$0.30 & 1800 & 12.73$\pm$0.36 & 2050 & 3.18$\pm$0.27 & 2300 \\

\enddata
\tablenotetext{a}{Fluxes in units of 10$^{-14}$ erg cm$^{-2}$ s$^{-1}$ and FWHM
in \kms. The subcripts n,b stand for narrow and broad components, respectively}
 
 \end{deluxetable}

\clearpage

 \begin{deluxetable}{lccc}
 \tablecaption{Measured and expected flux for the \feii\ lines \lb1840,1844\tablenotemark{a,b} \label{balance}}
 \tablewidth{0pt}
 \tablehead{
 \colhead{Galaxy} & \colhead{Observed\tablenotemark{c}} & \colhead{Expected} & \colhead{E(B-V)}}
 \startdata
 1H\,1934-063 &   13.2   &  158.0  &   0.61 \\
 Ark\,564     &   0.7    &  58.0   &   1.10 \\
 Mrk\,335     &   1.7    &  74.9   &   0.93 \\
 Mrk\,1044    &   0.6    &  57.0   &   1.12 \\
 \enddata
 \tablenotetext{a}{In units of 10$^{-14}$ erg cm$^{-2}$ s$^{-1}$}
 \tablenotetext{b}{The values of columns 2 and 3 correspond to the sum of the flux of
 the individual \feii\ features}
 \tablenotetext{c}{Upper limit}
 \end{deluxetable}

 \clearpage

 \begin{deluxetable}{lcccccccc}
 \tablecaption{\feii\ line ratios for the observed sample \label{feiirat}}
 \tabletypesize{\scriptsize}
 \tablewidth{0pt}
 \tablehead{
 \colhead{Ratio} & \colhead{1H\,1934-063} & \colhead{Ark\,564} &
 \colhead{Mrk\,335}  & \colhead{Mrk\,1044}  & \colhead{I\,Zw\,1\tablenotemark{1}}
 & \colhead{LkH$\alpha$\,101\tablenotemark{2}} & \colhead{MWC 349\tablenotemark{5}} &
 \colhead{R Mon\tablenotemark{5}}
 }
 \startdata
 \lb9997/\lb10171  & 5.32$\pm$1.78 & 4.53$\pm$0.85 & 5.66$\pm$0.90 & 6.18$\pm$2.00 & \nodata & 3.70 & 3.5  & \nodata \\
 \lb9997/\lb10501  & 1.72$\pm$0.20 & 1.32$\pm$0.11 & 2.38$\pm$0.31 & 1.17$\pm$0.29 & 2.22$\pm$0.29 & 1.11 & 1.6 & 1.5 \\
 \lb9997/\lb10863  & 1.07$\pm$0.10 & 1.74$\pm$0.16 & \nodata       & 1.04$\pm$0.15 & 0.90$\pm$0.15 & 1.26 & \nodata & 0.98 \\
 \lb9997/\lb11127  & 3.47$\pm$0.65 & 3.19$\pm$0.43 & 7.92$\pm$1.05 & 2.51     &   4.59$\pm$0.76    & 1.59 & 2.00 & 1.9 \\
 \lb11127/\lb10863 & 0.31$\pm$0.06 & 0.55$\pm$0.08 & \nodata       & \nodata     &   0.20$\pm$0.04 & 0.80 & \nodata & 0.55 \\
 \lb10501/\lb10863 & 0.62$\pm$0.07 & 1.31$\pm$0.14 & \nodata       & 0.89$\pm$0.24 & 0.41$\pm$0.07 & 1.14 & \nodata & 0.70 \\
 \lb10501/\lb11127 & 2.02$\pm$0.41 & 2.40$\pm$0.35 & 3.33$\pm$0.94 & 2.23     &  2.06$\pm$0.37     & 1.43 & 1.20  & 1.3  \\
 \feii\ \lb4750/H$\beta$ & 0.89\tablenotemark{3} & 0.81\tablenotemark{4} & 0.47\tablenotemark{4} & 0.91\tablenotemark{4} & 1.34\tablenotemark{4}
 & \nodata & \nodata & \nodata \\

 \enddata

 \tablerefs{
 (1) Rudy et al. 2000; (2) Rudy et al. 1991 (3) Rodr\'{\i}guez-Ardila,
 Pastoriza
 \& Donzelli 2000; (4) Joly 1991; (5) Kelly, Rieke \& Campbell 1994}

 \end{deluxetable}

 \end{document}